\documentstyle[12pt,aasms4,flushrt,psfig]{article}
\newcommand{\bm}[1]{{\mbox{\boldmath$#1$}}}
\lefthead{S.M. Kopeikin \& L.M. Ozernoy}
\righthead{ 
Post-Newtonian Theory for Precision Doppler Measurements}
\begin{document}

\title {
Post-Newtonian Theory for Precision Doppler Measurements\\ of Binary
Star Orbits 
}
\author{S. M. Kopeikin\altaffilmark{1,2}}
\affil{
Theoretisch-Physikalisches Institut, Friedrich-Schiller-Universit\"at, 07743 
Jena, Germany}
\and
\author{L. M. Ozernoy\altaffilmark{3,4}}
\affil{Physics \& Astronomy Department and Institute for Computational
Sciences \& Informatics,\\ 5C3, George Mason University, Fairfax, 
VA 22030-4444 }
\altaffiltext{1}{On leave from: ASC of P.N. Lebedev Phys. Institute, PRAO, 
Leninskii Prospect,
53, Moscow 117924, Russia}
\altaffiltext{2}{e-mail: smk@tpi.uni-jena.de }
\altaffiltext{3}{also Laboratory for Astronomy and Solar Physics, NASA, Goddard
Space Flight Center}
\altaffiltext{4}{e-mail: ozernoy@science.gmu.edu}
%\tableofcontents
%\newpage
\begin{abstract}
The determination of velocities of stars from precise Doppler measurements
is described here using relativistic theory of astronomical reference frames 
so as to determine the Keplerian and post-Keplerian parameters of binary 
systems. Seven reference frames are introduced: (i) proper frame of a particle
emitting light, (ii) the star-centered reference frame, (iii) barycentric
frame of the binary, (iv) barycentric frame of the Galaxy, (v) barycentric
frame of the Solar system, (vi) geocentric frame, and (vii) topocentric
frame of observer at the Earth. We apply successive Lorentz transformations
and the relativistic equation of light propagation to establish the exact
treatment of Doppler effect in binary systems both in special and general
relativity theories. As a result, the Doppler shift is a sum of (1) linear
in $c^{-1}$ terms, which include the ordinary Doppler effect and its
variation due to the secular radial acceleration of the binary with respect
to observer; (2) terms proportional to $c^{-2}$, which include the
contributions from the quadratic Doppler effect caused by the relative
motion of binary star with respect to the Solar system, motion of the
particle emitting light and diurnal rotational motion of observer, orbital
motion of the star around the binary's barycenter, and orbital motion of the
Earth; and (3) terms proportional to $c^{-2}$, which include the
contributions from redshifts due to gravitational fields of the star, star's
companion, Galaxy, Solar system, and the Earth. After parameterization of
the binary's orbit we find  that the presence of periodically changing terms 
in the Doppler schift enables us 
disentangling different terms and measuring, along with the well known
Keplerian parameters of the binary, four additional post-Keplerian 
parameters, which characterize: (i) the relativistic advance of the 
periastron; (ii) a combination of the quadratic Doppler and gravitational 
shifts associated with the orbital motion of the primary relative to the 
binary's barycentre and with the companion's gravitational field, 
respectively; (iii) the amplitude of the `gravitational lensing' contribution
to the Doppler shift; and (iv) the usual inclination angle of the 
binary's orbit, $i$ . We briefly discuss feasibility of practical implementation 
of these theoretical results, which crucially depends on further progress
in the technique of precision Doppler measurements.
\end {abstract}


%\tableofcontents
\newpage
\section{Introduction}

Binary stars represent perhaps the most valuable targets for stellar       
astrophysics. They have been a source of insight into the structure and
evolution of stars, theory of radiative transfer, stellar magneto- and
hydrodynamics, and the Newtonian theory of gravity, to mention just a few 
major topics (Shore 1994). After discovery of
the first binary pulsar 1913+16, these objects have become excellent
gravitational laboratories for testing general relativity 
%as one of the most fundamental theories among the natural sciences 
by using radio observations (Taylor 1992, 
1994). Nevertheless, optical observations of binary stars continue to remain 
one of the most important sources of getting qualitatively new astronomical 
information. The reason is that the amount of binary stars
observable at optical wavelenghts overwhelmingly exceeds the number of 
objects accessible for radio, X-ray, and/or $\gamma $-ray observations. 
Moreover, distribution of the
relative orientation of binary orbits ranging from the edge-on to the
face-on, allows one to obtain valuable information in studying different
stellar phenomena. Therefore, increasing accuracy of optical measurements of
binary stars is a real challenge for modern astronomy. This is why precision 
Doppler measurements of stellar spectra implemented recently for the 
search of extrasolar planets could open a new direction in binary
star research.

Traditional techniques in radial velocity measurements rarely achieve 
an accuracy better than about 200 m s$^{-1}.$ Given such an uncertainty, 
usually the {\it linear} Doppler effect, i.e. the term 
of order of only $v/c$ could be only measurable. Its measurement 
for the primary star of mass $m_s$ belonging to a binary system brings
five classical Keplerian parameters of the stellar orbit. They are: projected
semimajor axis $a_s,$ eccentricity $e,$ orbital period $P_b,$ longitude of
the periastron $\omega ,$ and the epoch $T_0$ of the initial periastron
passage. Combination of these parameters makes it possible to calculate the
mass function for the binary system: 
\begin{equation}
\label{iii}
f(m_s, m_c)=\frac{m_c^3\sin ^3i}{(m_s+m_c)^2}, 
\end{equation}
where $m_c$ is the companion mass, and $i$ is the inclination angle of the 
stellar orbit to the line of sight. Obviously, the information which
can be extracted from the linear Doppler shift alone is incomplete to 
determine {\it all} orbital parameters of the binary, including 
the masses of its constituent stars. Observations of additional relativistic 
effects are necessary. They
include: relativistic advance of the periastron, quadratic Doppler and
gravitational redshifts, monotonous decrease of orbital period due to the
emission of gravitational waves, and effects of deflection and retardation
of electromagnetic waves in the companion's gravitational field. All of
them have been observed in binary pulsars (Taylor 1992) but only
relativistic advance of periastron could be measured with the use of
conventional spectroscopic technique (Semeniuk \& Paczynski 1968, Guinan \&
Maloney 1985). 

Ozernoy (1997a, 1997b) pointed out that current accuracy of Precision Doppler
Measurements (PDMs) using the iodine-based Doppler technique (Valenti,
Butler \& Marcy 1995, Cochran 1996) is able to catch the second-order
in $v/c$ effects. Moreover, he also has shown, by taking into account 
special relativity alone, that binary stars offer
a unique opportunity to disentangle the linear and quadratic in $v/c$ terms 
and extract such an important parameter as inclination angle.

Even without addressing any concrete applications, a coherent, unambiguous 
interpretation of PDMs having an accuracy of $\sim$1 m s$^{-1}$ or better,
requires an adequate development of a relativistic theory.  By present, the
basic principles to construct such a theory have been well established. 
Recently, they have been worked out in a series of publications by Brumberg 
\& Kopeikin (1989a, 1989b, 1990) (BK approach) and Damour, Soffel \& Xu 
(1991, 1992, 1993) (DSX approach). The main idea is to introduce one global 
and several local coordinate charts in a gravitating system consisting of 
$N$ bodies to describe adequately the properties of space-time curvature both 
on the global scale
and locally in the vicinity of each body. The subsequent application of the
mathematical technique to match the asymptotic expansions of the metric tensor
in different coordinate systems allows one to obtain general relativistic
transformations between these systems, which generalize Lorentz 
transformations of special relativity. Practical conclusions of two approaches
are the same. In this paper, we use the BK approach 
to describe self-consistently the relativistic algorithm of PDMs.

The paper is organized as follows. In section 2, seven appropriate coordinate 
frames are introduced. Sec.~3 deals with the
Doppler effect in special relativity,  which can be
important for interpretation of PDMs in the situations, when effects of the
gravitational field are negligibly small. The post-Newtonian coordinate
transformations are sketched in section 4, along with the derivation of
the equation for propagation of photons in a gravitational field. 
The Doppler effect in
general relativity is explored in Section 5. Parameterization of the
Doppler effect and the explicit Doppler shift curve are given in section 6.
Section 7 outlines some observational implications 
of the theory. Finally, section 8 contains discussion and our conclusions.
The approach developed in this paper was earlier reported in Kopeikin
\& Ozernoy (1996).

\section{Coordinate Frames}

A rigorous mathematical treatment of precision Doppler observations of a 
binary star requires the use of seven (4-dimensional) reference frames
(RFs):

$(G)-$ barycentric reference frame of our Galaxy $(cT,\vec X)=(X^0,X^i);$

$(S)-$ the Solar system's barycentric reference frame $(ct,\vec
x)=(x^0,x^i); $

$(B)-$ the binary system's barycentric reference frame $(cs,\vec
z)=(z^0,z^i);$

$(C)-$ the star-centered reference frame\footnote{It is important to emphasize
here that $\lambda$ denotes coordinate time in the star-centered reference
frame and is not a wavelenght of photon.} $(c\lambda ,\vec \eta )=(\eta
^0,\eta ^i);$

$(E)-$ the Earth's (geocentric) reference frame $(cu,\vec w)=(w^0,w^i);$

$(T)-$ topocentric reference frame of terrestrial observer $(c\tau ,\vec \xi
)=(\xi ^0,\xi ^i);$

$(P)-$ proper reference frame of a particle emitting light $(c\upsilon ,\vec
\zeta )=(\zeta^0,\zeta ^i).$\noindent
\medskip
Here and hereinafter, the arrow above the letter denotes a spatial vector 
having three
coordinates; index ``0"  relates to time coordinate, and small latin
indices (such as $i=1,2,3$) represent the spatial coordinates.

The origin of each coordinate frame coincides with the center of mass
(barycenter) of the respected system of gravitating bodies. For
instance, the origin of the Solar system RF is at the center of mass of the
Solar system, the origin of the emitting star's RF is at the center of mass
of the star, and so on. The observer is regarded to be massless and placed
at the origin of the topocentric RF. We assume that emission of light is
produced by the atom placed at the origin of its own proper RF $(P)$.
Each RF has its own coordinate time. These times are related to each other by
means of relativistic time transformations (Brumberg \& Kopeikin 1989a,
1989b, 1990). It is worth noting that the coordinate time of the topocentric
RF coincides precisely with the proper time of the observer measured by the
atomic clocks, and the coordinate time of the RF of the atom emitting light
coincides with its proper time.
The barycentric RF of our Galaxy is considered to be asymptotically flat so
that it covers all space-time. All other coordinate frames are not
asymptotically flat, and they cover only restricted domains in space because 
of a non-zero space-time curvature. All coordinate systems 
are assumed to be nonrotating in the kinematical sense (Brumberg \& Kopeikin
1989a). It means that spatial axes of all RF's are aligned and anchored to
the outermost quasars whose proper motions are negligibly small.

To derive equations describing the Doppler effect at the post-Newtonian level 
of accuracy, we use the relativistic post-newtonian transformations between
the coordinate frames. They have been formulated by Kopeikin (1988) and
Brumberg \& Kopeikin (1989a, 1989b), and are discussed briefly in 
section 4. However, for pedagogical reasons, it is useful to consider the
Doppler effect first in the framework of special relativity. The special
relativistic approach is motivated by the fact that one can get the
{\it exact} treatment of the problem under consideration, which will provide 
a guide of consistency for general relativistic calculations. It should
be noted, however, that the special relativistic approach is not able to
take into account the relativistic effects asssociated with an accelerated 
motion of bodies as well as influence of gravitational field. Those efffects 
can be adequately considered only in framework of general relativity.

\section{Doppler effect in Special Relativity}

The Doppler effect in special relativity is usually considered only
for two RFs: one is assumed to be at rest, and the other moves
with respect to the first one with a constant velocity. Here we discuss a
more realistic situation when five coordinate systems $S,B,C,E$ and $T$, 
introduced in the previous section, need to be considered. This
situation is rather close to the real astronomical practices and it might be
applied to the interpretation of spectral observations of binary stars if
the influence of gravitational fields could be neglected. In this section, 
we abandon, for the sake of simplicity, the reference frames $G$ and $P$. The
reason is that, for the moment, we want to avoid the discussion of terms
caused by the motion of the Solar and binary systems about the center of our
Galaxy, as well as motions of the emitting particles with respect to the star.
Accounting for these effects will be done in later sections.

The RF $S$ is the basic one which is considered to be at rest. The origins of
the RFs $B$ and $C$ are moving with respect to $S$ with constant relative
velocities ${\bf V}_B,$ and ${\bf V}_C$, respectively$.$ The RF $C$
is supposed to move with respect to the $B$ with a constant relative
velocity ${\bf v}_C.$ Note that, in the relativistic approach, ${\bf v}_C\neq 
{\bf V}_C.$ The RFs $E$ and $T$ move with constant velocities ${\bf V}_E$
and ${\bf V}_T$ with respect to $S.$ The relative velocity of the reference
frame $T$ with respect to $E$ is ${\bf v}_T$ (again, ${\bf v}_T\neq {\bf V}%
_T $ $).$

Three different approaches for discussion of Doppler effect can be applied.
They are based, accordingly, on the techniques of relativistic frequency 
transformation
(Landau \& Lifshitz 1951), successive Lorentz transformations (Weinberg
1972), and time transformations along with the equation of light propagation
(Tolman 1934, Brumberg 1972).

\subsection{Frequency transformation technique}

Let $k^\alpha$ be the 4-vector of the electromagnetic wave propagating 
from the source of light to the observer. Here and hereinafter,
greek indices run from 0 to 3, i.e. $k^\alpha =(k^0,k^i).$ This is a null
vector in the flat space-time. Therefore, $k^0=2\pi \nu/c,$ and  
$k^i=-k^0n^i,$  where $\nu $ is the frequency of the electromagnetic wave and the unit
spatial vector $n^i$ is tangent to the trajectory of the light ray. 
For convenience, it is chosen to be directed from the observer toward the 
point of emission.
Let $u^\alpha=(u^0, u^i)$ be the vector of 4-velocity of a massive particle.
The time component is $u^0=1/\gamma ,$ and the spatial components are $%
u^i=u^0\beta ^i,$ where $\gamma =(1-\beta^2)^{1/2},$ is the (constant) 
Lorentz-factor,
$\beta ^i=v^i/c,$ and $v^i$ is a spatial velocity of the particle%
$.$ By contracting vectors $k^\alpha $ and $u^\alpha $, one forms a
scalar which is relativistically invariant as it is independent of the choice of
reference frame: 
\begin{equation}
\label{a0}u_\alpha k^\alpha =u^ik^i-u^0k^0\equiv{\rm invariant,} 
\end{equation}
where the repeated spatial indices mean a summation from 1 to 3.

Suppose that the source of light moves with respect to the observer with a
constant speed $v^i.$ The 4-velocity of the observer in its proper RF is 
defined
as $u^\alpha =(1,0,0,0),$ and the 4-velocity of the source of light is $%
u^\alpha =\gamma ^{-1}(1,\beta ^i).$ Let the frequency of the emitted
electromagnetic wave be $\nu _0,$ and the received frequency be $\nu .$
Then, by applying equation (\ref{a0}) to the two different RF's one gets
(boldface letters denote spatial vectors,
and the dot in between two spatial vectors stands for usual scalar product): 
\begin{equation}
\label{a1}
\frac{\nu _0}\nu =\frac{1+({\bm {\beta} {\bf \cdot n}})}{(1-\beta ^2)^{1/2}%
}\;, 
\end{equation}
which is a well-known result (Landau \& Lifshitz 1951) for the Doppler
shift of the frequency of  light emitted by a moving source and received by 
the observer at rest. Here the unit spatial vector ${\bf n}$ is
measured with respect to the observer's RF.

While applying this formula, we will use slightly different notations:
 $\nu _{*}$ for the frequency of the emitted light and  $\nu $ for the observed 
frequency, viz.,  $\nu _s,\nu _b,$, and $\nu _e$ for the frequencies of light 
observed in reference frames $S,B,$ and $E$, respectively. Let us
also introduce a fractional frequency 
shift function, $z=\frac{\nu _{*}}\nu -1$. Succesive application
of equation (\ref{a1}) yields 
\begin{equation}
\label{a2}1+z=\frac{\nu _{*}}{\nu _b}\frac{\nu _b}{\nu _s}\frac{\nu _s}{\nu
_e}\frac{\nu _e}\nu =\frac{1+({\bm {\beta} }_C{\bf \cdot N}_B)}{(1-\beta
_C^2)^{1/2}}~\frac{1+({\bm {\beta} }_B{\bf \cdot N})}{(1-\beta _B^2)^{1/2}}%
~\frac{(1-\beta _E^2)^{1/2}}{1+({\bm {\beta} }_E{\bf \cdot N})}~\frac{(1-\beta
_T^2)^{1/2}}{1+({\bm {\beta} }_T{\bf \cdot N}_E)}. 
\end{equation}
Here ${\bf N=(x_{*}-x)/\left| x_{*}-x\right| }$ is the spatial unit vector
tangent to the light ray and having components measured with respect
to the coordinate system $S$, and ${\bf N}_B{\bf =
(z_{*}-z)/| z_{*}-z| }$ and $
{\bf N}_E ={\bf (w_{*}-w)/| w_{*}-w| }$ represent the same tangent
vector with the components measured relative to the systems $B$ and $E$,
respectively. Equation (\ref{a2}) contains the dimensionless particle 
velocities ${\bm %
\beta }_B{\bf =V}_B{\bf /}c$, ${\bm {\beta} }_C{\bf =v}_C{\bf /}c$, ${\bm %
\beta }_E{\bf =V}_E{\bf /}c$, and ${\bm {\beta} }_T{\bf =v}_T{\bf /}c$.
Coordinates with the asterisk concern the point of emission of light,
and coordinates without asterisk are related to the point of observation. It is
worth noting that the components of vectors ${\bf N}_B$ and ${\bf N}_E$ do
not coincide with those of vector ${\bf N}$ because of the relativistic
aberration of light.

A remarkable feature of the formula (\ref{a2}) is that it represents
the Doppler effect as a product of four different multipliers. Each factor
describes transformation of frequency of light from one reference frame to
another.  It would be straightforward to generalize  this result for the
description of the Doppler effect in the event of as many reference frames
as necessary. Because of importance of equation (\ref{a2}), it is instructive 
to derive it by using different techniques and then to compare the results.

\subsection{Lorentz transformation technique}

Lorentz transformations from the reference frame $S$ to $B$ is described by
the matrix $\Lambda _\beta ^\alpha $ with the components (Weinberg 1972,
Brumberg 1972, 1991): 
\begin{equation}
\label{a3}
\begin{array}{c}
\begin{array}{ccc}
\Lambda _B^{00}=\gamma _B^{-1}, &  & \Lambda _B^{0i}=\Lambda _B^{i0}=-\gamma
_B^{-1}\beta _B^i, 
\end{array}
\\  
\\ 
\Lambda _B^{ij}=\delta ^{ij}+(\gamma _B^{-1}-1)\beta _B^{-2}\beta _B^i\beta
_B^j. 
\end{array}
\end{equation}
Similarly, the Lorentz transformation from the reference frame $B$ to $C$
is described by the matrix $\Lambda _C^{\alpha \beta }$ with the components: 
\begin{equation}
\label{a4}
\begin{array}{c}
\begin{array}{ccc}
\Lambda _C^{00}=\gamma _C^{-1}, &  & \Lambda _C^{0i}=\Lambda _C^{i0}=-\gamma
_C^{-1}\beta _C^i, 
\end{array}
\\  
\\ 
\Lambda _C^{ij}=\delta ^{ij}+(\gamma _C^{-1}-1)\beta _C^{-2}\beta _C^i\beta
_C^j. 
\end{array}
\end{equation}
The Lorentz transformations from the reference frame $S$ to $E,$ and from $E$
to $T$ are given respectively by the matrices $\Lambda _E^{\alpha \beta }$
and $\Lambda _T^{\alpha \beta }$with the components: 
\begin{equation}
\label{a5}
\begin{array}{c}
\begin{array}{ccc}
\Lambda _E^{00}=\gamma _E^{-1}\;, &  & \Lambda _E^{0i}=\Lambda _E^{i0}=-\gamma
_E^{-1}\beta _E^i\;, 
\end{array}
\\  
\\ 
\Lambda _E^{ij}=\delta ^{ij}+(\gamma _E^{-1}-1)\beta _E^{-2}\beta _E^i\beta
_E^j\;, 
\end{array}
\end{equation}
\vspace{ 1.5 cm} 
\begin{equation}
\label{a5a}
\begin{array}{c}
\begin{array}{ccc}
\Lambda _T^{00}=\gamma _T^{-1}\;, &  & \Lambda _T^{0i}=\Lambda _T^{i0}=
-\gamma_T^{-1}\beta _T^i\;, 
\end{array}
\\  
\\ 
\Lambda _T^{ij}=\delta ^{ij}+(\gamma _T^{-1}-1)\beta _T^{-2}\beta _T^i\beta
_T^j\;.\bigskip 
\end{array}
\end{equation}
The relationship between time components of a light ray's 4-vector  is given
by the successive Lorentz transformations (the repeated greek indices imply
 summation from 0 to 3): 
\begin{equation}
\label{a6}k_C^0=\Lambda _C^{0\beta }\Lambda _B^{\beta \gamma }k_S^\gamma\; , 
\end{equation}
\begin{equation}
\label{a7}k_T^0=\Lambda _T^{0\beta }\Lambda _E^{\beta \gamma }k_S^\gamma\; , 
\end{equation}
where $k_S^\alpha ,k_C^\alpha ,$ and $k_T^\alpha $ are components of the light 
ray vector referred to the RF's $S,C,$ and $T$, respectively$.$

By substituting the matrices of the Lorentz transformations into
equations (\ref{a6}),(\ref{a7}) 
and defining the null vector $k_S^\alpha =2\pi \nu c^{-1}(1,-N^i)$,
we obtain after straightforward calculations
the following result: 
\begin{eqnarray}
\label{a8}
1+z&=&\frac{(1-\beta _T^2)^{1/2}}{(1-\beta _C^2)^{1/2}}\;\frac{%
(1-\beta _E^2)^{1/2}}{(1-\beta _B^2)^{1/2}}\;\times\\\nonumber\\\nonumber
\mbox{}&&\times\;\frac{1+({\bm {\beta} }_B{\bf %
\cdot N})+\gamma _B({\bm {\beta} }_C{\bf \cdot N})+({\bm {\beta} }_C{\bm \cdot
{\beta} }_B)+(1-\gamma _B)\beta _C^{-2}({\bm {\beta} }_C{\cdot \bm{ \beta} }_B)(%
{\bm {\beta} }_B{\bf \cdot N})}{1+({\bm {\beta} }_E{\bf \cdot N})+\gamma_E
({\bm %
\beta}_T\cdot {\bf N})+({\bm {\beta} }_T{\cdot \bm{ \beta} }_E)+(1-\gamma
_E)\beta _T^{-2}({\bm {\beta} }_T{\cdot \bm{ \beta} }_E)({\bm {\beta} }_E{\bf %
\cdot N})}.
\end{eqnarray}
At first sight, it looks quite different compared to equation (\ref{a2}).
However, by making relativistic transformation of vectors ${\bf N}_E$ and 
${\bf N}_B$ to vector ${\bf N}$ in equation (\ref{a2}),
 one can readily show that both expressions are completely identical. 
 In the rest of thid section, we derive, for the 
reader's convenience,  the transformation law between vectors 
${\bf N}_E$ and ${\bf N}$. (The transformation
law between vectors ${\bf N}_B$ and ${\bf N}$ is obtained similarly by
replacing index $E$ for $B$ and coordinates $w^i$ for $z^i.$)

The transformation between spatial coordinates of RFs $S$ and $E$
is given by (Weinberg 1972, Brumberg 1972, 1991): 
\begin{equation}
\label{ap1}w^i=\Lambda _E^{ij}(x^j-V_E^jt)\;, 
\end{equation}
where the transformation matrix $\Lambda _E^{ij}$ is defined in equation
(\ref{a5}). In its explicit form, the transformation (\ref{ap1}) reads
\begin{equation}
\label{ap2}{\bf w}={\bf x-V}_Et+\left[ \frac 1{(1-V_E^2/c^2)^{1/2}}-1\right] 
\frac{{\bf V}_E\cdot ({\bf x-V}_Et)}{V_E^2}{\bf V}_E\;. 
\end{equation}
Let us express the coordinates of radius-vector ${\bf N}_E$ connecting
points of emission and observation in the RF $E$ through the
coordinates of vector ${\bf N}.$ From eq. (\ref{ap2}) and equation (\ref{d})
for light propagation one gets: 
\begin{equation}
\label{ap3}w_{*}^i-w^i=\left| {\bf x}_{*}-{\bf x}\right| \Lambda
_E^{ij}(N^j+\beta _E^j)\;, 
\end{equation}
and, as a consequence, 
\begin{equation}
\label{ap4}\left| {\bf w}_{*}-{\bf w}\right| =\left| {\bf x}_{*}-{\bf x}%
\right| \frac{1+({\bm {\beta} }_E\cdot {\bf N})}{(1-V_E^2/c^2)^{1/2}}\;. 
\end{equation}
Now it is easy to obtain the relationship between vectors ${\bf N}_E$ and $%
{\bf N}$, whisch is given by: 
\begin{equation}
\label{ap5}
N_E^i=\frac{(1-{\bm {\beta} }_E^2)^{1/2}}{1+({\bm {\beta }}_E{\bf %
\cdot N)}}\left[ N^i+\gamma _E^{-1}\beta _E^i+(\gamma _E^{-1}-1)\frac{({\bm 
{\beta} }_E{\bf \cdot N)}\beta _E^i}{\beta _E^2}\right] . 
\end{equation}
Inversely, vector $N^i$ is obtained from equation (\ref{ap5}) after
replacements $N^i\longrightarrow N_E^i$ ,$N_E^i\longrightarrow N^i$, and $\beta
_E^i\longrightarrow -\beta _E^i$:
\begin{equation}
\label{ap6}N^i=
\frac{(1-{\bm {\beta} }_E^2)^{1/2}}
{1-({\bm {\beta }}_E{\bf \cdot
N}_E)}\left[ N_E^i-\gamma _E^{-1}\beta _E^i+(\gamma _E^{-1}-1)\frac{
({\bm {\beta }}_E{\bf \cdot N}_E)\beta _E^i}{\beta _E^2}\right] . 
\end{equation}

The transformations (\ref{ap5}) and (\ref{ap6}) represent, in fact, general
expressions for the relativistic aberration of light rays. This can be seen
from the relativistic law of addition of velocities
(Weinberg 1972, Brumberg 1972, 1991). In case under consideration, it is
given by: 
\begin{equation}
\label{ap7}V^i=\frac{(1-{\bm {\beta} }_E^2)^{1/2}}{1+c^{-1}({\bm {\beta }}_E
{\bf \cdot v)}}\left[ v^i+\gamma _E^{-1}\beta _E^i+(\gamma _E^{-1}-1)\frac
{{\bm %
(\beta }_E{\bf \cdot v)}\beta _E^i}{\beta _E^2}\right] , 
\end{equation}
where $v^i$ and $V^i$ are the relative velocities of a particle with respect 
to RFs $E$ and $S$, respectively. For the light particle (photon)
these velocities are $v^i=-cN_E^i,$ and $V^i=-cN^i.$ Having
substituted them to eq. (\ref{ap7}), one obviously gets eq. (\ref{ap6}).

Finally, using equations (\ref{ap5}) - (\ref{ap7}), one obtains: 
\begin{equation}
\label{ap9}1+({\bf N}_E\cdot {\bm{ \beta} }_T)=\frac{1+({\bm {\beta} }_E{\bf %
\cdot N})+\gamma _E({\bm {\beta} }_T{\bf \cdot N})+({\bm {\beta} }_T \cdot
{\bm{\beta }}_E)+(1-\gamma _E)\beta _T^{-2}({\bm {\beta} }_T\cdot{\bm{\beta}}_E)
({\bm {\beta} }_E{\bf \cdot N})}{1+({\bf N}\cdot {\bm{\beta} }_E)}\;. 
\end{equation}
One can see from equation (\ref{ap9}) and a similar expression for $1+
({\bf N}_B{\cdot \bm{ \beta} }_C)$
that equations (\ref{a2}) and (\ref{a8}) are identical. The
advantage of eq. (\ref{a8}) over (\ref{a2}) is that
only one vector ${\bf N}$ enters eq. (\ref{a8}), instead of three vectors $%
{\bf N}_E{\bf ,N}_B{\bf ,}$ and ${\bf N}$ in eq. (\ref{a2}).

\subsection{Time transformation technique}

Previous techniques used to derive the Doppler equation have not taken into 
account an
essential fact of separation of the two events -- emission and observation of
light -- in space-time. In fact, we have implicitly assumed that the
null vector $k^\alpha $ is the same at the points of emission and
observation of light. However, this is only true for a very special case
of negligent gravitational field and propagation of light in vacuum. In
general, this conditions are not met. Therefore, a more advanced
technique is required to tackle the Doppler effect appropriately. Such a 
technique, based on the integration of the equation for light propagation 
from the point of emission
to the point of observation, establishes a relationship between
coordinates of the 4-vector of a photon at these two events. Transformation
laws of time scales between different RFs are to be taken into
account as well. This approach, being rather general and straightforward, 
can be applied to analyse any particular situation. In this section, 
we consider time transformation technique in special relativity only. 
Its application to observations of binary stars in the framework of general 
relativity will be discussed in later sections.

The equation of light propagation, in the absence of gravitational
field and interstellar medium, is quite simple: 
\begin{equation}
\label{a9}x^i=x_{*}^i+cN^i(t-t_{*}), 
\end{equation}
\begin{equation}
\label{d}t-t_{*}=\frac 1c\left| {\bf x_{*}}-{\bf x}\right| , 
\end{equation}
where $t^{*}$ is the instant of photon emission and $t$ is the instant of 
observation of the photon, both measured as coordinate time of RF
$S$, in which ${\bf x_{*}=x}(t_{*})$ is the point of emission, and ${\bf %
x=x}(t)$ is the point of observation. It is worth noting that, although 
the instants $t$ and $t_{*}$ belong to the same RF $S$, their increments 
$\Delta t$ and $\Delta t_{*}$ are different
because of a relative motion of the source of light and the observer.

One can see from equation (\ref{a9}) that when the influences of 
gravitational field and the medium are both negligent, the components of the 
vector $%
k^\alpha $ are constant everywhere on the light ray's trajectory. This makes
clear why we do not care about a point in space-time in
which earlier we calculated the Doppler shift. However, in a more general
situation, the equations (\ref{a9}), (\ref{d}) are not so simple, and this 
point has to be appropriately taken into consideration.

Doppler effect is described by the function $1+z=\frac{\nu _{*}}\nu ,$ where 
$\nu _{*}=1/\Delta \lambda _{*}$ is the frequency of the emitted light, and $%
\nu =1/\Delta \tau $ is the observed frequency. By taking
the time intervals to be infinitesimally small, we get a differential
formula: 
\begin{equation}
\label{e}1+z=\frac{d\tau }{du}\frac{du}{dt}\frac{dt}{dt_{*}}\frac{dt_{*}}{%
ds_{*}}\frac{ds_{*}}{d\lambda _{*}}, 
\end{equation}
which is nothing more but a simple rule for differentiation of an
hierarchical function $f(\lambda _{*})=\tau (u(t(t_{*}(s_{*}(\lambda
_{*}))))).$ This result demonstrates as well that the Doppler effect can 
be presented
as a product of several multiplyers. A difference between equations (\ref{e}) 
and (\ref{a2}) is that in (\ref{e}) we use coordinate times of the respected
RFs and distinguish explicitly the points of emission and observation of 
light. Meanwhile in equation (\ref{a2}) only proper frequencies of the
electromagnetic wave are considered. The advantage of eq. (\ref{e})
is that, for its derivation, one needs to know only relativistic
transformations between time scales, whereas transformation law between 
spatial coordinates is not required. As we shall see later on, this advantage 
is very helpful while tackling the Doppler effect in general relativity.

To calculate the time derivatives at the points of emission and observation,
one needs incorporating
time components of the Lorentz transformations (\ref{a3}) - (\ref
{a5a}) between different RFs. They are: 
\begin{equation}
\label{a}\tau =\frac{u-c^{-1}({\bm {\beta} }_T\cdot {\bf w})}{(1-\beta
_T^2)^{1/2}}\;, 
\end{equation}
\begin{equation}
\label{b}u=\frac{t-c^{-1}({\bm {\beta} }_E\cdot {\bf x})}{(1-\beta
_E^2)^{1/2}}\;,
\end{equation}
\begin{equation}
\label{c}s_{*}=\frac{t_{*}-c^{-1}({\bm {\beta} }_B\cdot {\bf x_{*}})}{(1-\beta
_B^2)^{1/2}}\;, 
\end{equation}
\begin{equation}
\label{c1}\lambda _{*}=\frac{s_{*}-c^{-1}({\bm {\beta} }_C\cdot {\bf z_{*}})}{%
(1-\beta _C^2)^{1/2}}\;. 
\end{equation}

Let us remind that observer is fixed with respect to the RF $T$%
, and the source of light is fixed with respect to $C.$ Therefore, ${\bf %
\beta }_T=c^{-1}{\bf v}_T=c^{-1}d{\bf w}/du,$ and ${\bm {\beta} }%
_C=c^{-1}{\bf v}_C=c^{-1}d{\bf z}_{*}/du.$ Velocities of the
observer and the source of light relative to the RF $S$ are $%
{\bf V}_T=d{\bf x}/dt$ and ${\bf V}_C=d{\bf x}_{*}/dt$,
respectively (it is important to note that ${\bf V}_T\neq {\bf v}_T$ and $%
{\bf V}_C\neq {\bf v}_C)$. Therefore one obtains from (\ref{a}%
) - (\ref{c1}) : 
\begin{equation}
\label{f}\frac{d\tau }{du}=(1-\beta _T^2)^{1/2}\;, 
\end{equation}
\begin{equation}
\label{g}\frac{du}{dt}=\frac{1-({\bm {\beta} }_E\cdot {\bf V}_T)}{(1-\beta
_E^2)^{1/2}}\;, 
\end{equation}
\begin{equation}
\label{h}\frac{dt_{*}}{ds_{*}}=\frac{(1-\beta _B^2)^{1/2}}{1-c^{-1}({\bm %
\beta }_B\cdot {\bf V}_C)}\;, 
\end{equation}
\begin{equation}
\label{h1}\frac{ds_{*}}{d\lambda _{*}}=(1-\beta _C^2)^{-1/2}\;. 
\end{equation}
In addition, differentiation of equation (\ref{d}) gives: 
\begin{equation}
\label{i}\frac{dt}{dt_{*}}=\frac{1+c^{-1}({\bf N}\cdot {\bf V}_C)}{1+c^{-1}(%
{\bf N}\cdot {\bf V}_T)}\;. 
\end{equation}
Substitution of expressions (\ref{f}) - (\ref{i}) into eq. (\ref{e})
gives for the Doppler shift: 
\begin{equation}
\label{k}
1+z=\frac{(1-\beta _T^2)^{1/2}}{(1-\beta _C^2)^{1/2}}\;\frac{%
(1-\beta _B^2)^{1/2}}{(1-\beta _E^2)^{1/2}}\;\frac{1-c^{-1}({\bm {\beta} }%
_E\cdot {\bf V}_T)}{1-c^{-1}({\bm {\beta} }_B\cdot {\bf V}_C)}\;\frac{1+c^{-1}(%
{\bf N}\cdot {\bf V}_C)}{1+c^{-1}({\bf N}\cdot {\bf V}_T)}\;. 
\end{equation}
This equation does not coincide apparently neither with (\ref{a2}), nor 
with (%
\ref{a8}). Nethertheless, taking into account relativistic transformations
between vectors ${\bf N}_E$, ${\bf N}_T$, and ${\bf N}$ as well as the law
of addition of spatial velocities one can readily show that all three 
expressions for the Doppler effect are identical.
Indeed, with the use of equation (\ref{ap7}) it follows that 
\begin{equation}
\label{ap8}\frac{1-c^{-1}({\bm {\beta} }_E\cdot {\bf V}_T)}{(1-\beta
_E^2)^{1/2}}=\frac{(1-\beta _E^2)^{1/2}}{1+({\bm {\beta} }_E{\cdot \bm{ \beta} }%
_T)}\;, 
\end{equation}
and 
\begin{equation}
\label{ap10}1+c^{-1}({\bf N\cdot V}_T)=\frac{1+({\bm {\beta} }_E{\bf \cdot N}%
)+\gamma _E({\bm {\beta} }_T{\bf \cdot N})+({\bm {\beta} }_T\cdot {\bm{ \beta}} 
_E)+(1-\gamma _E)\beta _T^{-2}({\bm {\beta} }_T\cdot {\bm{ \beta} }_E)({\bm {
\beta }}_E{\bf \cdot N})}{1+({\bm {\beta} }_E\cdot {\bm{ \beta} }_T)}\;. 
\end{equation}
These relationships, along with ones obtained from (\ref{ap8}) - (\ref{ap10})
after replacement of indices $T$ to $C$ and $E$ to $B$, allow to see that 
expression (\ref{k}) for the Doppler effect coincides with (\ref{a8}) and,
consequently, with (\ref{a2}).

This completes the derivation of the exact equations for the Doppler effect 
in special relativity. These equations could be expanded into powers of 
$1/c$ to get an approximate solution. Unfortunately, we would not be able 
to apply directly those expressions to real astronomical practices since 
gravitational fields of the Solar system, the binary 
system, and the Galaxy give contributions comparable
with the special relativistic quadratic Doppler shift.
Thus, in order to explore the Doppler effect in general relativity, 
it is important to elaborate approximative analytical methods .
To tackle this problem, we apply the relativistic theory
of astronomical reference frames developed by Kopeikin (1988) and Brumberg \&
Kopeikin (1989a, 1989b).

\section{Coordinate transformations in General Relativity}

Transformation laws between reference frames in general relativity 
generalize the Lorentz transformations of special relativity.
They can be derived in two steps. First of all, the explicit form of metric
tensor in different RFs are obtained by solving the Einstein
equations with relevant boundary conditions. Then, the general relativistic
transformations between the RFs are derived using the method of matched 
asymptotic technique. A clear and simple introduction to this technique
is given in Brumberg \& Kopeikin (1990). Here we give the transformation
laws in the form which is suitable for discussion of the Doppler effect 
with a more than sufficient accuracy.
To derive the Doppler shift, we apply the technique based on time
transformations (Sec.~3.3). Thus, there is no need for development of
relativistic part of space-time transformation between spatial coordinates,
which will be given hereinafter only in the Newtonian approximation.

\subsection{Transformation between topocentric and geocentric reference frames}

This transformation law is given by:

\begin{equation}
\label{1}\tau =u-\frac 1{c^2}\left[ \int \left( \frac 12v_{{\rm T}}^2+\Phi _{%
{\rm T}}\right) du+v_{{\rm T}}^k(w^k-w_{{\rm T}}^k)\right] +O(c^{-4}), 
\end{equation}
\begin{equation}
\label{2}\xi ^i=w^i-w_{{\rm T}}^i+O(c^{-2}), 
\end{equation}
where $w_{{\rm T}}^k(u)$ and $v_{{\rm T}}^k(u)=dw_{{\rm T}}^k/du$ are
geocentric spatial coordinates (RF $E)$ and velocity of the observer,
respectively; and $\Phi _{{\rm T}}$ is the geopotential at the observer's
location site. It is worth noting that the quantity $\left(\frac 
12v_{{\rm T}}^2+\Phi _{{\rm T}}\right)$ is constant on the geoid surface.
In eq. (\ref{1}), the tidal gravitational potential of external bodies is not
included since it is negligibly small. Geocentric coordinates and observer's  
velocity both depend on time. Once the observer (spectrograph) is at the
surface of the Earth, its $w_{{\rm T}}^k$ and $v_{{\rm T}}^k$ are precisely
calculated using the data of the International Earth Rotation Service
(IERS). If the observer is on board of a satellite, its motion can be derived
using the satellite monitoring service.

\subsection{Transformation between geocentric and solar barycentric reference
frames}

This transformation is found in the form:
\begin{equation}
\label{3}u=t-\frac 1{c^2}\left[ \int \left( \frac 12v_{{\rm E}}^2+U_{{\rm E}%
}\right) dt+v_{{\rm E}}^k(x^k-x_{{\rm E}}^k)\right] +O(c^{-4})\;, 
\end{equation}
\begin{equation}
\label{4}w^i=x^i-x_{{\rm E}}^i+O(c^{-2})\;, 
\end{equation}
where $x_{{\rm E}}^k(t)$ and $v_{{\rm E}}^k(t)=dx_{{\rm E}}^k/dt$ are
respectively the spatial coordinates (RF $S)$ and velocity of the geocentre 
relative to the barycenter of the Solar system; and $U_{{\rm E}}$ is the
gravitational potential of the Solar system at the geocenter. If the
external (with respect to the Earth) bodies of the Solar system are
approximated by massive point particles, then 
\begin{equation}
\label{4a}U_{{\rm E}}=\sum_{k=1}^N\frac{Gm_k}{r_k}, 
\end{equation}
where $m_k$ is mass of the body $k$; $r_k$ is the distance from the body $k$
to the geocentre, and the sum is taken over all the external bodies of the 
Solar
system. The potential $\Phi _{{\rm T}}$ is not included in $U_{{\rm E}}$, in
accordance with general principles of construction of relativistic theory of
astronomical reference frames (Kopeikin 1988, Brumberg \& Kopeikin 1989a,
Brumberg \& Kopeikin 1989b). The tidal gravitational potentials of
the bodies external with respect to the Solar system are not included either, 
because they are too small to be important in the calculations of the Doppler 
effect. Barycentric coordinates and velocities of the Earth and other bodies 
of the Solar system can be calculated using the contemporary numerical 
theories of their motions (Standish 1982, 1993).

\subsection{Transformation between the solar and galactic reference frames}

This transformation law reads:
\begin{equation}
\label{5}t=T-\frac 1{c^2}\left[ \int \left( \frac 12V_{{\rm S}}^2+W_{{\rm S}%
}\right) dT+V_{{\rm S}}^k(X^k-X_{{\rm S}}^k)\right] +O(c^{-4}), 
\end{equation}
\begin{equation}
\label{6}x^i=X^i-X_{{\rm S}}^i+O(c^{-2}), 
\end{equation}
where $X_{{\rm S}}^k(T)$ and $V_{{\rm S}}^k(T)=dX_{{\rm S}}^k/dT$ are
spatial coordinates and velocity of the barycentre of the Solar system with
respect to the barycentre of our Galaxy; $W_{{\rm S}}$ is the gravitational
potential of the Galaxy at the barycentre of the Solar system (the
potentials $\Phi _{{\rm T}}$ and $U_{{\rm E}}$ should not be included). The
galactic coordinates, velocity of the Solar system, and gravitational
potential of the Galaxy at the Solar system barycentre are all not well known
quantities so far. To measure them more accurately would be one of many 
practical implications of precision Doppler measurements of stars.

\subsection{Transformation between the binary and galactic reference frames}

This transformation is similar to eq. (\ref{5}) and is given by:
\begin{equation}
\label{7}s=T-\frac 1{c^2}\left[ \int \left( \frac 12V_{{\rm B}}^2+W_{{\rm B}%
}\right) dT+V_{{\rm B}}^k(X^k-X_{{\rm B}}^k)\right] +O(c^{-4}), 
\end{equation}
\begin{equation}
\label{8}x^i=X^i-X_{{\rm B}}^i+O(c^{-2}), 
\end{equation}
where $X_{{\rm B}}^k(T)$ and $V_{{\rm B}}^k(T)=dX_{{\rm B}}^k/dT$ are
spatial coordinates and velocity of the barycentre of the binary system 
relative to the barycentre of our Galaxy, respectively; $W_{{\rm B}}$ is the
gravitational potential of the Galaxy at the barycentre of the binary system
(gravitational potential of the binary system should not be included).

\subsection{Transformation between the stellar and binary reference frames}

This transformation is similar to eq. (\ref{3}) and has the form:
\begin{equation}
\label{9}\lambda =s-\frac 1{c^2}\left[ \int \left( \frac 12v_{{\rm C}}^2+U_{%
{\rm C}}\right) ds+v_{{\rm C}}^k(z^k-z_{{\rm C}}^k)\right] +O(c^{-4}), 
\end{equation}
\begin{equation}
\label{10}\eta ^i=z^i-z_{{\rm C}}^i+O(c^{-2}), 
\end{equation}
where $z_{{\rm C}}^k(s)$ and $v_{{\rm C}}^k(T)=dz_{{\rm C}}^k/ds$ are
spatial coordinates and velocity of the primary star relative to the 
barycentre
of the binary system; and $U_{{\rm C}}$ is the gravitational potential of the
companion star. Gravitational potential of the primary star should not be
included in $U_{{\rm C}}$ for the same reason why the potential $U_{{\rm E}}$
does not include geopotential $\Phi _{{\rm T}}$. The potential $U_{{\rm C}}$
is given by: 
\begin{equation}
\label{10a}U_{{\rm C}}=\frac{Gm_c}r, 
\end{equation}
where $m_c$ is the companion mass, and $r$ is the distance between the
two stars in the binary.

\subsection{Transformation between the proper frame of an emitting atom and 
stellar reference frame}

This transformation law is given by:
\begin{equation}
\label{1aa}\upsilon =\lambda -\frac 1{c^2}\left[ \int \left( \frac 12v_{{\rm %
P}}^2+\Phi _{{\rm P}}\right) d\lambda +v_{{\rm P}}^k(\eta ^k-\eta _{{\rm P}%
}^k)\right] +O(c^{-4}), 
\end{equation}
\begin{equation}
\label{2aa}\zeta ^i=\eta ^i-\eta _{{\rm P}}^i+O(c^{-2}), 
\end{equation}
where $\eta _{{\rm P}}^k(\lambda )$ and $v_{{\rm P}}^k(\lambda )=d\eta _{%
{\rm T}}^k/d\lambda $ are spatial coordinates and velocity of an emitting
atom relative to the star, respectively; $\Phi _{{\rm P}}$ is gravitational
potential of the star at the point of the atom's location.
Obviously, coordinates and velocity of a single emitting atom cannot be 
determined
since the integral flux of the stellar radiation is only observed. Motion of
the atom and gravitational potential of the star both causs the broadening 
of spectral lines in the stellar spectrum. This unfortunately complicates 
the precise measurement of the Doppler shift. In order to simplify discussion 
of this problem as much as possible, we assume here that $v_{{\rm P}}^k$, 
an average thermal velocity of atoms, is constant in time, and 
$\Phi_{{\rm P}}$, gravitational potential of the star at the altitude of 
the spectral line formation, is also a constant.

\subsection{Time transformation between instants of emission and observation}

Time transformation between instants of light  emission and observation is
obtained from the solution of equation for light propagation in vacuum, which
is described by the equation of isotropic geodesic line (Weinberg 1972).
Solution of this equation has a simple form in the galactic reference
frame $G$ so that the time interval between the instants of light emission,
$T_{*}$, and  observation, $T$ $(T>T_{*})$, is given by$:$ 
\begin{equation}
\label{11}T-T_{*}=\frac 1c\left| {\bf X}_{*}-{\bf X}\right| +\Delta _S(T,T%
{\bf _{*})}, 
\end{equation}
where $X_{*}^i$ are the galactic coordinates of the emitting atom at the
instant of emission, and $X^i$ are the galactic coordinates of the observer
at the instant of light observation. Relativistic correction $\Delta _S$
is  of order of $O(c^{-3}).$ It describes the Shapiro time delay (Shapiro 1964) 
in the gravitational field: 
\begin{equation}
\label{11aa}
\Delta _S=\sum_{a}\frac{2Gm_a}{c^3}\ln \frac{|{\bf X%
}_{*}-{\bf X}_a|+|{\bf X}-{\bf X}_a|+|{\bf X}_{*}-{\bf X}|}{|{\bf X}_{*}-%
{\bf X}_a|+|{\bf X}-{\bf X}_a|-|{\bf X}_{*}-{\bf X}|}, 
\end{equation}
where subscript $a$ stands for a body $a$ that deflects light rays, 
$X_a^i$ are its spatial coordinates
taken at the moment $T_a$ of the closest approach of the photon to the body
(Klioner \& Kopeikin 1992, Kopeikin et al. 1998). It can be shown (Brumberg 1972) that the main 
term in the Shapiro delay depends logarithmically upon $d$, the impact 
parameter of the light ray (for more detail see also the paper (Kopeikin 1997)).
The contribution to the Doppler shift
caused by the Shapiro delay is proportional to $(v/c)(r_g/d)$, where $v$
is the characteristic relative velocity, and $r_g=2GM/c^2$ is the
gravitational radius of the deflector. This estimate makes it obvious that 
the contribution (\ref{11aa}) to the Doppler shift produced by the Shapiro 
delay can be only substantial
in the nearly edge-on binary systems containing invisible relativistic
companion -- a neutron star or a black hole.
Functions ${\bf X}(T)$ and ${\bf X}_{*}(T_{*})$ can be decomposed into a sum
of vectors 
\begin{equation}
\label{11a}{\bf X}(T)={\bf X}_{{\rm S}}(T)+{\bf x}_{{\rm E}}(T)+{\bf w}_{%
{\rm T}}(T)+O(c^{-2}), 
\end{equation}
\begin{equation}
\label{11b}{\bf X}_{*}(T_{*})={\bf X}_{{\rm B}}(T_{*})+{\bf z}_{{\rm C}%
}(T_{*})+{\bm {\eta} }_{{\rm P}}(T_{*})+O(c^{-2}), 
\end{equation}
where the relativistic terms come from the
relativistic part of the transformation of spatial coordinates. 
In subsequent calculations of the Doppler shift in general relativity,
we will use eq. (\ref{11}) coupled with these expansions.

\section{Doppler effect in General Relativity}
\subsection{General equation}
Frequency of the emitted light is related to the proper time of the emitting 
atom as $%
\nu _{*}=1/\triangle \upsilon _{*},$ where $\triangle \upsilon _{*}$ is the
period of the emitted electromagnetic wave. Frequency of the observed light 
is $\nu =1/\triangle \tau ,$ where $\triangle \tau $ is the period of the
received electromagnetic wave. The Doppler shift $z$ in frequency is 
calculated as a product of the appropriate time derivatives: 
\begin{equation}
\label{13}1+z=\frac{d\tau }{d\upsilon _{*}}=\frac{d\tau }{du}\frac{du}{dt}%
\frac{dt}{dT}\frac{dT}{dT_{*}}\frac{dT_{*}}{ds_{*}}\frac{ds_{*}}{d\lambda
_{*}}\frac{d\lambda _{*}}{d\upsilon _{*}}, 
\end{equation}
where $d\tau /du,$ $du/dt,$ $dt/dT$ are taken at the instant of observation;
 $dT_{*}/ds_{*},$ $ds_{*}/d\lambda _{*},$ $d\lambda _{*}/d\upsilon
_{*}$ are taken at the instant of emission; and $dT/dT_{*}$ is
calculated by finding the differential of the left and right hand
sides of equation (\ref{11}) for propagation of light. Thus, equation (\ref
{13}) is not just the usual time derivative taken at the same point of
space-time. On the contrary,  this is a two-point function that relates 
two events separated in space and time and connected by an isotropic 
worldline.
\subsection{Expansion into a series in $1/c$}
Time derivatives at the point of observation are obtained by direct
differentiation of equations (\ref{1}), (\ref{3}), and (\ref{5}), which 
describe
relativistic transformations between different time scales in the Solar
system. All these derivatives are taken at the point of observation:
\begin{equation}
\label{13a}\frac{d\tau }{du}=1-\frac 1{c^2}\left[ \frac 12v_{{\rm T}}^2+\Phi
_{{\rm T}}({\bf w}_{{\rm T}})\right] +O(c^{-4}), 
\end{equation}
\begin{equation}
\label{13b}\frac{du}{dt}=1-\frac 1{c^2}\left[ \frac 12v_{{\rm E}}^2+U_{{\rm E%
}}({\bf x}_{{\rm E}})\right] +\frac 1{c^2}a_{{\rm E}}^k(x^k-x_{{\rm E}%
}^k)+O(c^{-4}), 
\end{equation}
\begin{equation}
\label{13c}\frac{dt}{dT}=1-\frac 1{c^2}\left[ \frac 12V_{{\rm S}}^2+W_{{\rm S%
}}({\bf X}_{{\rm S}})\right] +\frac 1{c^2}\dot V_{{\rm S}}^k(X^k-X_{{\rm S}%
}^k)+O(c^{-4}). 
\end{equation}
Here $a_{{\rm E}}^k=dv_{{\rm E}}^k/dt$ is the acceleration of the geocentre
relative to the barycentre of the Solar system, and $\dot V_{{\rm S}%
}^k=dV_{{\rm S}}^k/dT$ is the acceleration of the barycentre of the Solar
system with respect to the barycentre of our Galaxy.

Calculations of time derivatives at the point of emission can be done with 
the help of equations (\ref{7}), (\ref{9}), and (\ref{1aa}):
\begin{equation}
\label{13aa}\frac{d\lambda _{*}}{d\upsilon _{*}}=1+\frac 1{c^2}\left[ \frac
12v_{{\rm P}}^2+\Phi _{{\rm P}}({\bf \eta }_{{\rm P}})\right] +O(c^{-4}), 
\end{equation}
\begin{equation}
\label{13d}\frac{ds_{*}}{d\lambda _{*}}=1+\frac 1{c^2}\left[ \frac 12v_{{\rm %
C}}^2+U_{{\rm C}}({\bf z}_{{\rm C}})\right] -\frac 1{c^2}a_{{\rm C}%
}^k(z_{*}^k-z_{{\rm C}}^k)+O(c^{-4}), 
\end{equation}
\begin{equation}
\label{13e}\frac{dT_{*}}{ds_{*}}=1+\frac 1{c^2}\left[ \frac 12V_{{\rm B}%
}^2+W_{{\rm B}}({\bf X}_{{\rm B}})\right] -\frac 1{c^2}\dot V_{{\rm B}%
}^k(X_{*}^k-X_{{\rm B}}^k)+O(c^{-4}). 
\end{equation}
Here $a_{{\rm C}}^k=dv_{{\rm C}}/dt$ is the acceleration of the emitting
star with respect to the binary system barycentre, and $\dot V_{{\rm B}}^k
=dV_{{\rm B}}^k/dT_{*}$ is the acceleration of the barycentre of the binary
system with respect to the barycentre of our Galaxy (all quantities are
calculated at the point of emission).

The function $dT_{*}/dT$ depends on two instants of time, {\it viz.}, 
emission and observation of light. We have found that it is more convenient 
to transform $dT_{*}/dT$
 to the instant of emission alone so that to express the final result 
through the
instantaneous relative velocity of the barycentre of the binary with respect
to the barycentre of the Solar system. It enables us to exclude from the 
final equation
for the Doppler shift the poorly known velocities of the binary 
and the Solar
systems with respect to the centre of mass of our Galaxy. To complete this
procedure, we introduce the notations as follows:
\begin{itemize}
\item
$R^i\equiv x_{{\rm B}}^i(t_{*})=X_{{\rm B}}^i(T_{*})-X_{{\rm S}%
}^i(T_{*})+O(c^{-2})$ is the relative distance between
the Solar system and binary barycentres taken at the instant of
emission;
\item
$K^i=R^i/R$ is the unit vector directed toward to the barycentre of the
binary (this vector slowly changes due to proper motion ${\bm {\mu} }$;
\item
$v^i=dR^i/dt_{*}$ is the relative velocity of the binary's  barycentre
relative to the barycentre of the Solar system, taken at the moment
of emission;
\item
$v_R^i=({\bf K\cdot v})K^i$ is the radial velocity of the binary's
barycentre;
\item
$v_T^i=\left[ {\bf K\times }\left[ {\bf v\times K}\right] \right] ^i={\bm %
{\mu} }R$ is the transverse velocity of the binary's barycentre.
\end{itemize}

The two-point time derivative $dT_{*}/dT$ can be found by
means of calculation of differential of equations (\ref{11}), (\ref{11a}),
and (\ref{11b}). This results in: 
\begin{equation}
\label{13f}
\frac{dT}{dT_{*}}=\frac{1+c^{-1}({\bf N\cdot V}_{*})+
c^{-3}F_{*}}{1+c^{-1}({\bf N\cdot V})+c^{-3}F}, 
\end{equation}
\begin{equation}
\label{13ff}F_{*}=2G\sum_{a}M_a\left[ \frac{({\bf N\cdot V_{*})-%
}({\bf n_{*}\cdot V}_{*}{\bf )+(n_{*}\cdot V_a)}}{R_{*a}+R_a-D}+\frac{({\bf %
N\cdot V_{*})+}({\bf n_{*}\cdot V}_{*}{\bf )-(n_{*}\cdot V}_a{\bf )}}{%
R_{*a}+R_a+D}\right]\;,\bigskip 
\end{equation}
\begin{equation}
\label{13gg}F=2G\sum_{a}M_a\left[ \frac{({\bf N\cdot V)+}({\bf %
n\cdot V)-(n\cdot V}_a{\bf )}}{R_{*a}+R_a-D}+\frac{({\bf N\cdot V)-}({\bf %
n\cdot V)+(n\cdot V}_a{\bf )}}{R_{*a}+R_a+D}\right]\;,
\end{equation}
where ${\bf V,}$ ${\bf V}_{*}$, and ${\bf V}_a$ are galactic velocities of
the observer, source of light, and deflecting body $a$, respectively; $%
R_{*a}=\left| {\bf X}_{*}(T_{*})-{\bf X}_a(T_a)\right|$; $R_a=\left| {\bf X}%
(T)-{\bf X}_a(T_a)\right|$; $D=\left| {\bf X}(T)-{\bf X}_a(T_a)\right|$;
and the unit vectors ${\bf N,}$ ${\bf n_{*}}$, and ${\bf n}$ are defined as: 
\begin{equation}
\label{13g}{\bf N}=\frac{{\bf X}_{*}(T_{*})-{\bf X}(T)}{\left| {\bf X}(T)-%
{\bf X}_a(T_a)\right| }\;, 
\end{equation}
\begin{equation}
\label{13g1}{\bf n_{*}}=\frac{{\bf X}_{*}(T_{*})-{\bf X}_a(T_a)}{\left| {\bf %
X}_{*}(T_{*})-{\bf X}_a(T_a)\right| }\;, 
\end{equation}
\begin{equation}
\label{13g2}{\bf n}=\frac{{\bf X}(T)-{\bf X}_a(T_a)}{\left| {\bf X}(T)-{\bf X%
}_a(T_a)\right| }\;. 
\end{equation}
Furthermore, equation (\ref{13f}) is expanded into the powers of $c^{-1},$ 
and it can be simplified using the relationships: 
\begin{equation}
\label{13g3}R_{*a}+R_a-D=\frac{d^2}2\frac{R_{*a}+R_a}{R_{*a}R_a}+O(d^4)\;, 
\end{equation}
\begin{equation}
\label{13g4}R_{*a}+R_a+D=2(R_{*a}+R_a)+O(d^2)\;, 
\end{equation}
\begin{equation}
\label{13g5}{\bf n_{*}}={\bf N}+\frac{{\bm {\xi} }}{R_{*a}}-\frac 12{\bf N}%
\left( \frac d{R_{*a}}\right) ^2+O(d^3)\;, 
\end{equation}
\begin{equation}
\label{13g6}{\bf n}=-{\bf N}+\frac{{\bm {\xi} }}{R_a}+\frac 12{\bf N}\left(
\frac d{R_a}\right) ^2+O(d^3)\;, 
\end{equation}
where ${\bm {\xi} }=[{\bf N}\times [{\bf R}_{*a}\times {\bf N}]]=-[{\bf N}%
\times [{\bf R}_a\times {\bf N}]]$ is the vector of impact parameter $d$
pointing from the deflector to the light ray: $d=|{\bm {\xi} }|\ll \min
(R_a,R_{*a}).$

If we only  take into account the Shapiro effect in the binary
system, then $R_a\gg R_{*a}$ and equation (\ref{13f}) takes the form: 
\begin{eqnarray}
\label{13fg}
\frac{dT}{dT_{*}}=1+\frac 1c\left( {\bf N\cdot V_{*}}\right) -\frac
1c\left( {\bf N\cdot V}\right) {\bf +}  
+\frac 1{c^2}\left( 
{\bf N\cdot V}\right) ^2-\frac 1{c^2}\left( {\bf N\cdot V_{*}}\right)
\left( {\bf N\cdot V}\right) + \nonumber\\\mbox{}&&  \\\nonumber 
+\frac{2GM_c}{c^3}\left( -2\frac{{\bm {\xi} }}{d^2}+\frac{{\bf N}}{R_{*c}}%
\right) ({\bf V}_{*}-{\bf V}_c)\;. 
\end{eqnarray}
We have neglected in (\ref{13fg}) all terms of the order of  $V^3/c^3$ 
and higher, as well as those `mixed' terms
from the differentiation of $\Delta _S$, which are of the order of $%
(R_{*a}/R_a)(r_g/d)(V/c),$ $(d/R_{*a})^2(r_g/d)(V/c),$ and so on, where $V$
is the characteristic relative velocity of the companion relative to the
primary, $r_g=2GM_c/c^2$ is gravitational radius of companion, and $d$ is
the impact parameter of the light ray.

To continue, we expand the function ${\bf X}_{{\rm S}}(T)$ in eq. 
(\ref{11a}) into the time series near the instant $T_{*}$ : 
\begin{equation}
\label{13h}{\bf X}(T)={\bf X}_{{\rm S}}(T_{*})+{\bf V}_{{\rm S}%
}(T_{*})(T-T_{*})+{\bf x}_{{\rm E}}(T)+{\bf w}_{{\rm T}}(T)+O\left[
(T-T_{*})^2\right] , 
\end{equation}
and, instead of $T-T_{*}$, we substitute the r.h.s. of equation (\ref{11}).
The result is used to expand the unit vector ${\bf N}$ from eq. (\ref{13g}) 
into the powers of parallactic terms of the order of $x_{{\rm E}%
}/R,$ $z_{{\rm C}}/R,$ and so on. One gets: 
\begin{equation}
\label{13i}
{\bf N}={\bf K}+{\bm {\pi} _{{\rm C}}-{\bm{\pi}} _{{\rm E}}-}\frac
1c\left[ {\bf K\times }\left[ {\bf V_{{\rm S}}}(T{\bf _{*})\times K}\right]
\right] +O(\epsilon ^{-2})+O(\epsilon ^{-1}\pi )+O(\pi ^2). 
\end{equation}
Here the term depending on the velocity ${\bf V_{{\rm S}}}$ describes the
secular aberration. The binary orbital parallax ${\bm {\pi} _{{\rm C}}}$
(caused by the orbital motion of the star) as well as the annual parallax ${\bf %
\pi _{{\rm E}}}$ (caused by the orbital motion of the Earth) are given by: 
\begin{equation}
\label{13j}
{\bm {\pi} _{{\rm C}}=}\frac 1R\left[ {\bf K\times }\left[ {\bf z_{%
{\rm C}}\times K}\right] \right] , 
\end{equation}
\begin{equation}
\label{13k}
{\bm {\pi} _{{\rm E}}=}\frac 1R\left[ {\bf K\times }\left[ {\bf x_{%
{\rm E}}\times K}\right] \right] . 
\end{equation}
In equation (\ref{13i}), we also neglect, as currently unmeasurable, all 
terms of the order of $w_{{\rm T}}/R$ and $\eta _{{\rm P}}/R$.
Similarly to decomposition of functions ${\bf X}(T)$ and ${\bf X}_{*}(T_{*})
$, given by eqs. (\ref{11a}) and (\ref{11b}), decomposition of velocities 
${\bf V}(T)$ and ${\bf V}_{*}(T_{*})$ can be done. Then it is 
straightforward to show that 
\begin{equation}
\label{13m}{\bf V}_{*}(T_{*})-{\bf V}(T)={\bf v+v}_{{\rm P}}(T_{*})+{\bf v_{%
{\rm C}}}(T{\bf _{*})-v_{{\rm E}}(}T{\bf )-v_{{\rm T}}(}T{\bf )-}\frac 1c%
{\bf \dot V}_{{\rm S}}(T_{*})R+O(c^{-2}). 
\end{equation}
\subsection{The Doppler shift}
After substitution of the intermediate equations of the previous subsection 
into the basic equation (\ref{13}), the final result for the Doppler shift 
takes the form: 
\begin{equation}
\label{14}z(t)=z_C+z_{R\odot }+z_{E\odot }-z_R-z_E-z_S-z_M, 
\end{equation}
where partial contributions are given by:\bigskip\ 
\begin{equation}
\label{14a}
\begin{array}{c}
z_C=\frac 1cv_R+\frac 1{c^2}\left( \frac 12v_R^2+\frac 12\mu ^2R^2+W_{
{\rm S}}-W_{{\rm B}}-({\bf K\cdot \dot V}_{{\rm S}})R\right) + \\  \\ 
+\frac 1{c^2}\left( \frac 12v_{{\rm P}}^2+\Phi _{{\rm P}}\right) -\frac
1{c^2}\left( \frac 12v_{{\rm T}}^2+\Phi _{{\rm T}}\right) , 
\end{array}
\end{equation}
\bigskip\ 
\begin{equation}
\label{14b}z_{R\odot }=-\frac 1c({\bf K\cdot v}_{{\rm T}})-\frac
1c(1+c^{-1}v_R)({\bf K\cdot v}_{{\rm E}}), 
\end{equation}
\bigskip\ 
\begin{equation}
\label{14c}
z_{E\odot }=-\frac 1{c^2}\left[ \frac{v_{{\rm E}}^2}2+U_{{\rm E}%
}-({\bf K\cdot v}_{{\rm E}})^2\right] _{\mbox{constant}}-\frac 1{c^2}\left[ 
\frac{v_{{\rm E}}^2}2+U_{{\rm E}}-({\bf K\cdot v}_{{\rm E}})^2\right] _{%
\mbox{periodic}}, 
\end{equation}
\bigskip\ 
\begin{equation}
\label{14d}z_R=-\frac 1c({\bf K\cdot v}_{{\rm P}})-\frac 1c(1+c^{-1}v_R)(%
{\bf K\cdot v}_{{\rm C}}), 
\end{equation}
\bigskip\ 
\begin{equation}
\label{14e}z_E=-\frac 1{c^2}\left( \frac{v_{{\rm C}}^2}2+U_{{\rm C}%
}\right) _{\mbox{constant}}-\frac 1{c^2}\left( \frac{v_{{\rm C}}^2}2+U_{%
{\rm C}}\right) _{\mbox{periodic}}, 
\end{equation}
\bigskip\ 
\begin{equation}
\label{14f}z_S=\frac d{ds}\frac{2Gm_c}{c^3}\ln [r_R-({\bf K}\cdot {\bf r}%
_R)], 
\end{equation}
\bigskip\ 
\begin{equation}
\label{14g}z_M=\frac 1{c^2}({\bf K\cdot v}_{{\rm C}})({\bf K\cdot v}_{{\rm E%
}})-\frac R{c^2}({\bm {\mu} \cdot v}_{{\rm C}}). 
\end{equation}
Here ${\bm {\mu} =\dot k=V}_T/R$ is the vector of proper motion of the
binary's barycentre; ${\bf r}_R$ is the radius-vector of the primary star 
relative to its companion; $r_R=|{\bf r}_R|$; and we neglect all terms 
of the order of $10^{-10}$ and higher. The nature of the partial components
in eq. (76) is as follows:

The term $z_C$ contains a linear Doppler shift caused by the radial 
velocity of the emitting particle. It also includes both the quadratic 
Doppler and the gravitational shifts caused by the relative motion of the 
binary and the gravitational potential of our Galaxy, respectively. 
Contribution from $\left(\frac 12v_{{\rm T}}^2+\Phi _{{\rm T}}\right)$
causes broadening spectral lines in the primary's spectrum. The geopotential
term $\left(\frac 12v_{{\rm P}}^2+\Phi _{{\rm P}}\right)$ is constant in time,
and all temporal
variations of $z_C$ are expected to be caused by a radial acceleration of 
the binary and/or its proper motion.

The term $z_{R\odot }$ describes a linear Doppler shift caused by the
rotational motion of the terrestrial observer with velocity 
${\bf v}_{{\rm T}}$ and the orbital motion of the Earth's centre of mass 
with velocity ${\bf v}_{{\rm E}}.$ This term includes 
the radial component of the binary's relative velocity.

The term $z_{E\odot }$ includes a sum of quadratic Doppler  and  
gravitational shifts caused, respectively,
by the orbital motion of the Earth relative to the barycentre of the
Solar system {\it and} the gravitational fields of the Sun and planets.

The term $z_R$ describes a linear Doppler shift caused by the radial 
velocity of source of light ${\bf v}_{{\rm P}}$ relative to the star's 
centre and the radial component of the orbital velocity of the star's centre 
of mass ${\bf v}_{{\rm C}}.$ This term, like $z_{R\odot }$, includes the 
relative radial velocity of the binary.

The term $z_E$ is a sum of quadratic Doppler and gravitational shifts 
caused by the orbital motion of the primary star with respect to the 
barycentre of the binary {\it and} by the gravitational field of the 
companion.

The term $z_S$ represents a Doppler shift caused by the Shapiro delay in 
propagation of light in the companion's gravitational field. 
This effect can only be detected in the
nearly edge-on bynary systems. Its magnitude is generally negligibly small.

The term $z_M$ describes a Doppler shift caused by the effect of coupling of
motions of the Earth and the primary star.

\section{The explicit Doppler shift curve}
\subsection{The necessity of parameterization}
Equation (\ref{14}) as such cannot be used for reduction of observational 
data. It should be re-written in a way which would clearly pinpoint the 
measurable parameters. 
We have also to assign the proper instant of time (``exposure mid-time'' $t$) 
to any particular observation of stellar spectrum (Cochran 1996). Moreover, 
the exposure mid-time should be properly referred to the instant of light 
emission. The determination of the exposure mid-time is a rather difficult 
technical problem
and we do not discuss it here (see Cochran 1996). As for the relationship
between the exposure mid-time and the instant of light emission, it follows
from the equation of light propagation (\ref{11}) and has the well-known
form extensively used, e.g., in pulsar timing data reduction programs
(Taylor \& Weisberg 1989, Doroshenko \& Kopeikin 1995):
\begin{equation}
\label{15aa}(1+z_C)\lambda _{*}=t-t_0+\Delta _{R\odot }+\Delta _{E\odot
}+\Delta _R+\Delta _E+\Delta _S.
\end{equation}
Here $t_0$ is the initial epoch of observations; $\Delta _{R\odot },\Delta
_{E\odot },\Delta _R,\Delta _E,\Delta _S$ are respectively the R\"omer and 
Einstein delays in the Solar system; and $\Delta _R,\Delta _E,\Delta _S$ are, 
accordingly, the R\"omer, Einstein,
and Shapiro delays in the binary system. Their explicit expressions can be
found in Damour \& Taylor (1992), Taylor \& Weisberg (1989), and Doroshenko 
\& Kopeikin (1990, 1995).

\subsection{Convenient vectors for tracking the binary system} 
Let us introduce a triad of the unit vectors $({\bf I}_0,{\bf J}_0,{\bf K})$
attached to the barycentre of the binary system (see Fig.1).
%in the same manner as described in (Damour \& Taylor, 1992, Fig.1).
The vector ${\bf K}$ 
is directed from the Solar system barycentre toward that of the binary 
system, and
vectors ${\bf I}_0,{\bf J}_0$ are in the plane of the sky with ${\bf I}_0$
directed to the east, and ${\bf J}_0$ to the north celestial pole. Two other
sets of the unit vectors, $({\bf I},{\bf J},{\bf K})$ and $({\bf i},
{\bf j},{\bf k})$ are also introduced, which are related to $({\bf I}_0,
{\bf J}_0,{\bf K})$ by means of two spatial rotations (Damour \& Deruelle, 
1986b): 
\begin{equation}
\label{15}
\begin{array}{c}
{\bf I}=\cos \Omega\;{\bf I}_0+\sin \Omega \;{\bf J}_0\;, \\   
{\bf J}=-\sin \Omega\; {\bf I}_0+\cos \Omega \;{\bf J}_0\;,\\ {\bf K}={\bf K}_0\;,
\end{array}
\qquad 
\begin{array}{c}
{\bf i}={\bf I}\;, \\ {\bf j}=\cos i\;{\bf J}+\sin i\;{\bf K}\;, 
\\ {\bf k}=-\sin i\;%
{\bf J}+\cos i\;{\bf K}\;. 
\end{array}
\end{equation}
In the above transformations, the angles $\Omega $ $(0\leq $$\Omega <2\pi )$
and $i$ $(0\leq i<\pi )$ designate the longitude of the ascending node 
of the primary's orbit and the inclination of the orbit to the plane of
the sky, respectively. Vector ${\bf I}$ is directed to the ascending node of 
the binary's orbit, and vectors (${\bf i,j}$) lie in the orbital plane in 
the sense of orbital motion.

Vector ${\bf K}$ slowly changes due to a proper motion of the binary 
\begin{equation}
\label{15a}{\bf K}={\bf K}_0+{\bm {\mu} }(t-t_0), 
\end{equation}
where $t$ is the current time, and $t_0$ is the initial epoch of
observations. Therefore, the relative velocity of the binary's barycentre 
with respect to the Solar system is given by: 
\begin{equation}
\label{16}{\bf V}=R(\mu _\alpha {\bf I}_0+\mu _\delta {\bf J}_0)+v_R{\bf K}%
_0+\dot {\bf v}_R(t-t_0), 
\end{equation}
where $R$ is the distance between the binary and the Solar systems; $v_R$ 
is the relative radial velocity ($v_R=\dot R)$ at the initial epoch 
$t_0$; $\dot v_R$ is the radial acceleration; $\mu _\alpha $ and 
$\mu _\delta $ are the respective components of the proper motion of the 
star in the sky.
\subsection{Relativistic terms in the Doppler shift}
Relativistic perturbations of the orbit of a binary system are described in
Klioner \& Kopeikin (1994). Using the Damour-Deruelle relativistic
parameterization of the orbital motion (Damour \& Deruelle 1985, see also
Klioner \& Kopeikin 1994), we get with the necessary accuracy: 
\begin{equation}
\label{17}\frac 1c({\bf K}_0{\bf \cdot v}_{{\rm C}})=K_s\left[ \cos (\omega
+A)+e\cos \omega \right] +O(c^{-3}), 
\end{equation}
\begin{equation}
\label{17aaa}K_s=nx_s(1-e^2)^{-1/2}, 
\end{equation}
where $x_s=a_s\sin i/c$ is the projection of the semimajor axis $a_s$ of the
primary's orbit onto the line of sight; and $n=2\pi /P_b$ is angular
frequency of the orbital motion ($P_b$ being the orbital period) given by
\begin{equation}
\label{17sss}n=\left( \frac{GM}{a_R^3}\right) ^{1/2}\left[ 1+\left( \frac{%
m_pm_c}{M^2}-9\right) \frac{GM}{2a_Rc^2}\right].
\end{equation}
Here $m_s$ and $m_c$ are masses of the primary star and its companion, 
respectively, $M=m_s+m_c,$ $a_R=a_s(m_s+m_c)/m_c+O(c^{-2})$ is the semimajor 
axis of 
the primary's relative orbit in harmonic coordinates 
(Damour \& Deruelle 1986b), and $e$ is the 
eccentricity of this orbit. 
The angle $\omega $ in eq. (\ref{17}) is the longitude of periastron, which 
includes a contribution of its relativistic advance: 
\begin{equation}
\label{17a}\omega =\omega _0+kA_e\;, 
\end{equation}
where $\omega _0$ is the position of the periastron at the initial epoch,
and $k$ is the post-Keplerian parameter of relativistic advance of the
periastron (Robertson 1938, Damour \& Sch\"afer 1985, Kopeikin \& Potapov 1994):
\begin{equation}
\label{17b}k=\frac{3G}{c^2}\frac{m_s+m_c}{a_R(1-e^2)}+O(c^{-4})\;. 
\end{equation}
As an example, for a binary system with the parameters $m_s=3M_{\odot },$ $%
m_c=1.4M_{\odot },$ and $e=0.4,$  the magnitude of the relativistic advance
is about $0^{\circ }.1$ $yr^{-1},$ $1.2^{\prime \prime }$ $yr^{-1},$ and $%
0.004^{\prime \prime }$ $yr^{-1}$ for the orbits which semimajor axes are
$10^{12}$ cm, 10$^{13}$ cm, and 10$^{14}$ cm, respectively.

The angle $A_e$ entering eq. (\ref{17a}) is the eccentric anomaly, which 
is related to the time through the true anomaly $U$ and the third Kepler's 
law: 
\begin{equation}
\label{17cc}A_e=2\arctan \left[ \left( \frac{1+e}{1-e}\right) ^{1/2}\tan
\frac u2\right] , 
\end{equation}
\begin{equation}
\label{19}U-e\sin U=n\lambda +\sigma , 
\end{equation}
where $\sigma $ is the (constant) orbital phase at the epoch of the first
passage of the periastron.
Additional Newtonian perturbations of the binary orbit (whenever they are 
observationally important) may be included into equation
(\ref{17}) using the usual approach based on the orbital osculating elements
(e.g. Shore 1992, p.34).

Coupling of orbital and proper motions of the binary gives the term 
\begin{equation}
\label{20}
\frac 1c( 
{\bm {\mu} \cdot v}_{{\rm C}})=  
-\frac{K_s}{\sin i}\left[ (\mu _\alpha \cos \Omega +\mu _\delta \sin
\Omega )S(u)+\cos i(\mu _\alpha \sin \Omega -\mu _\delta \cos \Omega
)C(U)\right] , 
\end{equation}
where $\Omega $ is the longitude of the ascending node of the orbit, and
functions $C(u)$ and $S(u)$ are given by
\begin{equation}
\label{21}C(U)=\cos (\omega +A_e)+e\cos \omega \;, 
\end{equation}
\begin{equation}
\label{22}S(U)=\sin (\omega +A_e)+e\sin \omega =-\frac{dC(u)}{d\omega }.
\end{equation}
The quadratic Doppler effect plus gravitational shift of the frequency in the
companion's gravitational field are given by
\begin{equation}
\label{26}
\frac{v_{{\rm C}}^2}2+U_{{\rm C}}
=-\frac 12\frac{Gm_c[2m_s+m_c(1-e^2)]}{a_R(1-e^2)(m_s+m_c)}-\frac{%
Gm_c[m_s+2m_c]}{a_R(m_s+m_c)}\frac e{1-e^2}\cos A_e\;. 
\end{equation}
\bigskip\ Comparision of equations (\ref{26}) and (\ref{14e}) yields 
\begin{equation}
\label{26a}
\left( z_E\right) _{\mbox{constant}}=\frac{Gm_c^2}{2c^2a_R\left(
m_s+m_c\right) }+\frac \Upsilon e\;,\bigskip 
\end{equation}
%\medskip\ 
\begin{equation}
\label{26b}\left( z_E\right) _{\mbox{periodic}}=\Upsilon \cos A_e\;, 
\end{equation}
where a new relativistic post-Keplerian parameter $\Upsilon $ is  given by
\begin{equation}
\label{26c}\Upsilon =\frac{Gm_c[m_s+2m_c]}{c^2a_R(m_s+m_c)}\frac e{1-e^2}. 
\end{equation}
For the parameters of binary systems given below eq. (\ref{17b}), 
the magnitude of $\Upsilon $ is about $1.3\cdot
10^{-7},1.3\cdot 10^{-8},$ and $1.3\cdot 10^{-9}$, respectively.

Finally, for the ``gravitational lens" term $z_S$ we obtain 
\begin{equation}
\label{27a}z_S=\frac{\Im \left\{e\sin A_e-\sin i\left[ \cos (\omega
+A_e)+e\cos \omega \right]\right\} }{1-e\cos u-\sin i\left[ \sin \omega (\cos
u-e)+(1-e^2)^{1/2}\cos \omega \sin u\right] }, 
\end{equation}
where the third relativistic post-Keplerian parameter $\Im $ is defined as
\begin{equation}
\label{27b}\Im =\frac{2Gm_c}{c^3}\frac n{(1-e^2)^{1/2}}. 
\end{equation}
It is worth noting that the inclination angle $i$ defines the shape of the 
function (\ref{27a}). Therefore, it can be considered as the forth 
post-Keplerian parameter (Taylor 1992).

The effect of gravitational lensing is, under usual circumstances, rather 
small and it will be problematic to measure it. For instance, if the binary 
system consists of a main sequence star $m_s=3M_{\odot }$ and a relativistic 
companion $m_c=1.4M_{\odot }$ on the orbit having the relative semi-major 
axis $%
a_R=10^{12} $ cm and eccentricity $e=0$, the magnitude of $z_S$ is only $%
1.1\cdot 10^{-9}, $ $2.4\cdot 10^{-9},$ and $7.7\cdot 10^{-9}$ for the
orbital inclinations $\sin i=0.95,$ $0.99$, and $0.999$, respectively.

\section{Implications of the Doppler shift curve}

The set of equations derived in the previous section makes it possible 
to analyse PDM observations of binary stars on a quantitative basis. In this 
section, we explore how to disentangle various effects entering the basic
equation (76) for the Doppler shift and extract measurable parameters.

\subsection{Effects of Earth rotation and orbital motion}
%: $z_{R\odot},z_{E\odot }$}

In order to extract a pure effect caused by the primary's motions, the
Earth rotation and orbital motion have to be subtracted form the total
Doppler shift. It can be
easily done using machine readable data on the Earth rotation parameters
(IERS Annual Report), the Earth  spatial coordinates and the geocentre's   
velocity 
(Standish 1982, 1993) as well as position and proper motion of the binary
star taken from the astrometric catalogue. Since this paper only deals 
with principal topics, we do not develop here an exact
technical framework for calculating $z_{R\odot },z_{E\odot }$ and
consider them in the following as well predictable functions of time.

\subsection{Effects of constant part of gravitational field and relative
motion between the Solar system and the binary star}

The effect of relative motion between the Solar system and the binary star,
has the main contribution of the order of $O(c^{-1})$, is associated with 
the radial velocity of the binary's barycentre  and its radial acceleration. 
Terms of the order of $O(c^{-2})$ include the transversal velocity component 
$\mu R$ squared and the acceleration of the Solar system's barycentre
relative to the barycentre of the Galaxy $\dot V_{{\rm S}}.$ In addition, 
the constant parts of the gravitational fields of our Galaxy, the Solar
system, the binary system, 
%star, 
and the geopotential as well as the quadratic Doppler shift caused by the 
orbital motion of the Earth and the primary star, all contribute at the
level of $O(c^{-2}).$ The geopotential, the quadratic Doppler shift caused by 
the orbital motion of the Earth, and the constant part of the gravitational 
field of the Solar system on the Earth's orbit all can be calculated based on 
the gravimetric data and modern ephemerides with an accuracy which high 
enough to exclude those terms from the function $z.$ The remaining terms can 
be used to extract an additional information on the distribution of 
gravitational field in our Galaxy.

\subsection{Effects of the orbital motion of the primary star and proper 
motion of the binary system}
%: $z_R$}

The classical Doppler effect associated with the orbital motion of a binary 
system is well known. It allows to measure five Keplerian parameters: 
(i) $n$, the
frequency of the orbital motion; (ii) $\sigma$, the initial orbital phase; 
(iii) $e$, eccentricity; (iv) $\omega_0$, initial position of the periastron; 
and (v) $x_s=a_s\sin i/c$, the projected semimajor axis of the primary's orbit.
%and (vi) $t_0$, initial epoch of observation. 
Among these five parameters, two parameters, $n$ and $x$, while  
combined together, make it possible to determine the mass function 
of the binary, $f(m_s,m_c)$. In the binary with an invisible (compact) 
companion, the knowledge of mass function sets up an upper limit
to the mass of the companion. In the event when the Doppler
shift curve of the companion is also observable, this limit can be put even
tighter.

Measurements of relativistic effects in the orbital motion can provide a 
unique tool to determine the masses of stars in the binary without bias. 
In binary pulsars, such a
procedure is widely used to determine the masses of neutron stars
(Taylor 1992). 

The proper motion of the binary leads to a gradual secular change in the
observable orbital elements $x_s$ and $\omega .$ This situation is quite
similar to that in the pulsar timing (Kopeikin 1994, 1996; Arzoumanian
et al. 1996).
Indeed, one can see from equations (\ref{20}) - (\ref{22}) that, because
of the smallness of proper motion, the term $\frac 1c({\bm {\mu} \cdot v}_{%
{\rm C}})(t-t_0)$ can be entirely absorbed into $\frac 1c({\bf K}_0%
{\bf \cdot v}_{{\rm C}})$ by means of redefinition of parameters $x_s$ and 
$\omega$. As a result, the observable values $x_s^{obs}$ and $\omega ^{obs}$
are shifted from their physically meaningful values $x_s$ and $\omega $ by 
\begin{equation}
\label{d1}x_s^{obs}=x_s+\delta x_s,\quad \omega ^{obs}=\omega +\delta \omega, 
\end{equation}
where 
\begin{equation}
\label{e11}\delta x_s=x_s\cot i\left( -\mu _\alpha \sin \Omega +\mu _\delta
\cos \Omega \right) (t-t_0), 
\end{equation}
and 
\begin{equation}
\label{f1}\delta \omega =\csc i\left( \mu _\alpha \cos \Omega +\mu _\delta
\sin \Omega \right) (t-t_0). 
\end{equation}

It is important to emphasize that the parameter $x_s$ changes because of 
secular variation of the inclination angle $i$ due to proper motion. 
Meanwhile the semimajor axis $a_s$ remains constant, because proper motion 
does not cause any dynamical force acting on the orbital plane. Hence, 
equation (\ref{e11}) can be re-written in a form similar to (\ref{f1}): 
\begin{equation}
\label{e12}\delta i=\left( -\mu _\alpha \sin \Omega +\mu _\delta \cos \Omega
\right) (t-t_0). 
\end{equation}

The increments $\delta x_s$ and $\delta \omega $ depend on time linearly and
can appear in observations as small secular variations of the Keplerian
parameters $x_p$ and $\omega .$ Specifically, one gets: 
\begin{equation}
\label{g1}\delta \dot x_s=1.54\cdot 10^{-16}x_s\cot i\left( -\mu _\alpha
\sin \Omega +\mu _\delta \cos \Omega \right) [s\mbox{ }s^{-1}], 
\end{equation}
\begin{equation}
\label{h11}\delta \dot \omega =2.78\cdot 10^{-7}\csc i\left( \mu _\alpha
\cos \Omega +\mu _\delta \sin \Omega \right) [\deg \mbox{ }yr^{-1}], 
\end{equation}
where the values $x_s,\mu _\alpha ,$ and $\mu _\delta $ are expressed
 in seconds ($s$) and milliarcseconds per year (mas/yr), respectively.

It is worth noting that, for binary systems with a negligibly small
eccentricity, the effect of a secular variation in $\omega $ (along with
the relativistic advance of periastron) is absorbed by the re-definition of
the orbital frequency and therefore, in such systems, it is not observable at
all. Indeed, whenever the eccentricity is negligibly small,
the argument $\omega +A_e$ takes the form 
\begin{equation}
\label{h12}\omega +A_e=\omega _0+\frac{2\pi }{P_b^{obs}}(t-t_0), 
\end{equation}
where $P_b^{obs}$ is the observable value of the orbital period: 
\begin{equation}
\label{h2}P_b^{obs}=P_b\left[ 1-k-\frac{P_b}{2\pi }\csc i(\mu _\alpha \cos
\Omega +\mu _\delta \sin \Omega )\right] . 
\end{equation}
Here $P_b$ is the physical value of the orbital period and we neglected in (%
\ref{h2}) all terms nonlinear in small parameters.

If both the classical and relativistic perturbations of orbital motion are
negligibly small, then as can be seen from eqs. (\ref{g1}) and (\ref{h11}), 
the observable secular variations of $x_s$ and $\omega $ parameters could be 
used to determine both the ascending node of the binary's orbit and the 
inclination angle of the orbit, $i.$

Finally, it is worth noting that the term $\frac R{c^2}({\bm {\mu} \cdot v}_{%
{\rm C}})$ entering the function $z_M$ has the structure  similar to the term 
$\frac 1c({\bm {\mu} \cdot v}_{{\rm C}})(t-t_0)$. This does not include an
explicit dependence on time and, therefore, only leads to a constant shift
of the orbital parameters $x_p$ and $\omega $, which therefore cannot be 
determined.

\subsection{The post-Keplerian parameters: $k,\Upsilon ,\Im,$ and $\sin i$}

The post-Keplerian parameters $k,\Upsilon ,\Im$, and $\sin i$ can be measured 
in binary systems having relativistic orbits. Of the set of these 
parameters,
the parameter $\Upsilon $ contributes at the highest order $O(c^{-2})$. 
However, it can be only disentangled with difficulty from the classical
parameter $K_s\cos \omega $, 
similarly to what happens in binary pulsars (Brumberg {\it et al. }1975). 
The separation is possible only if the
relativistic advance of the periastron, $k$, is high enough to measure the
change in $K_s\cos \omega $.
It should be noted that the term $K_se\cos\omega $ that enters eq.~(88) 
is constant in the systems with negligibly small orbital
perturbations. It has a secular change as the parameter $\omega $ is not
constant.

The parameter $k$ contributes only at the level of $O(c^{-3})$ but in a
secular way. This makes it measurement both easy and accurate, which has
been done in a number of photometric and spectroscopic binaries (see, e.g., 
Shakura 1985, Khaliullin 1985). 
If parameters $%
\Upsilon $ and $k$ are both measurable, then along with the mass function 
this would allow to determine separately the masses of both stars and the
orbital inclination. If the classical perturbation of parameter 
$\omega $ (for instance, caused by the oblateness of stars)  is
also substantial, then observations of parameter $\Upsilon $  would allow to 
separate the relativistic
contribution to the advance of periastron from the classical one. This could
be used to infer the oblatenesses of the stars.

The two remaining parameters $\Im $ and $\sin i$ contribute, 
in a quasi-periodic way, at the level of 
$O(c^{-3})$. If it is done independently of the other 
parameters, $\Im $ and $\sin i$ can only be measured 
in the nearly edge-on binary systems via the
determination of the amplitude and the shape of function $z_S$, in a manner
similar to the pulsar timing (Taylor 1992)$.$ The range parameter $\Im $
defines the amplitude of $z_C,$ and $\sin i$ characterizes its shape. Once 
the parameters $\Im $ and $\sin i$ are determined, the masses of both stars 
can be obtained with the use of the mass function.

\section{Conclusions and Discussion}
For the reader's convenience, we summarize here the main conclusions of
this paper, along with the references to relevant equations.

1. As discussed in Introduction, Precision Doppler Measurements (PDMs),
which provide accuracy better than a few meters per second, measure more 
than just the {\it radial} component of the velocity  -- they also catch a
contribution of the {\it transverse} component, i.e. the terms of the second
order in $v/c$. A source with perioically changing velocity components,
such as a binary star, allows a disentangling different velocities and 
extracting additional (post-keplerian) parameters of the binary. To this end, 
a detailed relativistic theory of the Doppler shift is required.

2. In special relativity, the Doppler shift of a spectral frequency from a 
binary is given by equation (4) or by equivalent expressions (11) and
(32). The calculated Doppler shift includes the contributions from:
(i) the motion of the binary's barycentre in the Galaxy; (ii) the motion 
of the primary star in the binary; (iii) the motion of the Solar system 
barycentre in the Galaxy; and (iv) the Earth motion relative to the Solar 
system's barycentre. 

3. In general relativity, accounting for additional effects is necessary,
which includes, among others, gravitational field in the binary and 
acceleration of the primary star relative to the binary's barycentre.
The total Doppler shift is given by equation (76), which partial components
are presented by equations (77)--(83), with detailed comments about 
the physical meaning of each term given at the end of Sec.~5.

4. Presence of periodically changing terms in equation (76) enables us 
disentangling different terms and measuring, along with the well known
Keplerian parameters of the binary, four additional post-Keplerian parameters 
($k,\Upsilon ,\Im,$ and $\sin i$) as well. The first three of them are
given by equations (92), (101), and (103), respectively. The $k$ parameter 
characterizes the relativistic advance of the periastron; $\Upsilon$ 
characterizes the quadratic Doppler and gravitational shifts associated
with the orbital motion of the primary relative to the binary's barycentre
and with the companion's gravitational field, respectively;
$\Im$ characterizes the amplitude of the `gravitational lensing' contribution
to the Doppler shift given by equation (102); and $i$ is the usual 
inclination angle of the binary's
orbit (the value of $i$ defines the shape of the `gravitational lensing'
term).

5. The post-Keplerian parameters $k,\Upsilon ,\Im,$ and $\sin i$ can be          
measured in binary systems, which are sufficiently close so as to make the
relativistic effects measurable (on the other hand, as discussed below, 
for too close binaries with very short periods and therefore with very 
high orbital velocities there is a potential problem with the determination 
of the exposure mid-time while performing PDMs).

Feasibility of practical implementation of the theory developed in this
paper, crucially depends on further progress in PDM techniques. Analyzing 
the spectrum of a star with respect to the `velocity metric' based on
the spectrum of the iodine absorption cell allows one to remove most of the
wavelength drifts of the spectrograph and the detector. It can provide an 
ultimate precision in measuring radial velocity of a star attaining 1 m s$%
^{-1} $ or better (Cochran 1996). It is interesting to compare this
precision with that in millisecond pulsars timing technique. Available data 
indicate that 
the limiting accuracy of pulsar timing measurements, $\delta t,$ with present
techniques is a few microseconds, or less, over timespan of many years 
(Taylor 1992).
Uncertainty in the velocity measurement is the product of radial
acceleration of the body under consideration and the error in timing
measurement. The star's radial acceleration with respect to the binary's
barycentre is proportional to $4\pi ^2a\sin i/P_b^2\sim 2\pi v/P_b,$ where $%
v $ is the radial component of orbital velocity, $a$ is the semimajor axis of 
the star's orbit, and $i$ is inclination  angle of the orbit.
Thus, for the typical binary pulsar PSR B1913+16, where $\delta
t=15$ $\mu$s, $P_b\sim 28000$ s, and the ratio $v/c$ $\sim $10$^{-3},$ one
gets the precision of timing velocity measurements $\delta v\sim 
$ $2\pi c(v/c)(\delta t/P_b)\sim 0.01$ m s$^{-1}.$ Thus, the PDMs of binary
stars based on iodine absorption cell technique cannot currently 
be considered competitive with timing technique for binary
pulsars as concerned the precise tests of relativistic gravity. Nevertheless,
relative accuracy of PDMs ($\delta v/c\sim 3\cdot 10^{-9}$) is comparable
with the magnitude of the second or (in some cases) even third order for
relativistic perturbations in binary systems. Therefore, implementation of 
relativistic theory for proper tackling with such precision measurements 
is inevitable.

The prospects for PDMs combined with the relativistic theory of Doppler
shift presented above seem to be even more bright if one takes into 
consideration that a new generation of instruments for PDMs is currently
emerging. Among them is the use of Fourier Transform Spectroscopy with
the Navy Prototype Optical Spectrometer being built by the U.S. Naval
Observatory (Armstrong et al. 1998), which is projected to achieve in a 
near future a velocity resolution of only 0.3 m/s (Hajian 1998).

An accuracy with which the orbital parameters of a binary can be determined
by applying the PDMs is restricted by an uncertainty related to the
determination of the exposure mid-time. It is interesting to compare PDMs 
with pulsar timing where one can actually measure the arrival time of signals 
and not only the dopler shift. This allows a phase-connected solution for
astrometric, spin, and orbital parameters of the pulsar which
contains a fit to integer numbers. That is the reason for the high
accuracy in pulsar timing experiment. More simply, the precision, with which 
the 
position of a pulsar on its orbit can be determined, is given by a 
relationship $\left( \delta r\right) _{PT}\sim c\cdot \left( \delta t
\right) _{PT}$ . Let us assume, for convenience, that PDM gives an infinite
precision in determination of stellar velocity  but the exposure mid-time is
determined with an error $\left( \delta t\right) _{PDM}$, which is about 1 s
(Cochran 1996)$.$ Then inaccuracy in determination of the star's orbital
position is given by $\left( \delta r\right) _{PDM}\sim v\cdot
\left( \delta t\right) _{PDM}$ $\sim 2\pi a\sin i/P_b\left( \delta t\right)
_{PDM}$. A comparision of the two  expressions above gives: 
\begin{equation}
\label{15abc}\frac{\left( \delta r\right) _{PDM}}{\left( \delta r\right)
_{PT}}=\frac{2\pi x}{P_b}\frac{\left( \delta t\right) _{PDM}}{\left( \delta
t\right) _{PT}}, 
\end{equation}
where $x=a\sin i/c.$ For binary systems having orbital parameters 
like those in PSR 1913+16 with $2\pi x/P_b\sim
v/c\sim 10^{-3}$ and $\left( \delta t\right) _{PT}\sim 15$ $\mu$s one has $%
\left( \delta r\right) _{PDM}=67\left( \delta r\right) _{PT}\sim 300$ km.
However, it is worth emphasizing that since the accuracy of timing
measurements has a low limit about 1 $\mu$s one can achieve a better
determination of the orbits of binary systems in which $v/c<10^{-6},$ 
i.e. for the systems with very long orbital periods. In this sense, PDMs are 
much better suited to search for planets orbiting the extrasolar stars than 
pulsar timing.

In this paper, we have only considered the post-Newtonian theory for PDMs
of binaries consisting of a primary (optical) star and an (invisible) compact
companion. Although in many binaries, especially spectroscopic ones, measuring 
the Doppler effect for both stars would provide additional possibilities
for disentangling the orbital parameters, this kind of binary seems
to be less appropriate for PDM measurements. Indeed, in close binaries, where
relativistic effects could be measurable with PDMs, those effects 
would be severely contaminated by tidal interaction of the 
components and stellar winds, whereas in wide binaries the post-Newtonian
terms are expected to be rather weak. 
   
It is worthwhile to remind that normal stars subjected to the PDMs, 
in most cases, cannot be 
considered as point masses, unlike neutron stars or black holes. Due to
this reason, the classical perturbations will presumably be the most important
sources of orbital parameters' variations  (Shore 1994). However, even in
the situation when the classical perturbations  dominate, measuring of
(or proper taking into account) the relativistic effects will serve as a tool
to better understand the nature of the process(es) responsible for the
orbital parameters' variations. In this respect, interesting observational 
targets for PDMs might be massive main-sequence stars with radio pulsars in 
binary systems like PSR B1259-63 (Wex et al. 1988) or PSR J0045-7319. Timing 
observations reveal
(Lai, Bildstein \& Kaspi 1995) that it is possible to measure orbital
evolution caused by hydrodynamical effects associated with the optical 
companion but induced by the tidal gravitational field of the companion
(the pulsar) and/or intrinsic rotational motion of the primary 
star. PDMs of such a system, to be done complementary to
timing observations, would allow to measure tidal oscillations of the optical
star and therefore to essentially improve our knowledge about these systems. 
Another interesting application of PDMs could be an anomalous 
binary systems like DI Her, where some discrepancy was found  between the 
prediction based on general relativity and the
observed motion of the periastron (Guinan \& Maloney 1985). It
also can be explained by the dynamical influence of tidally-generated
oscillations on the orbital motion of stars but accuracy of the employed
observational technique was not good enough to test this hypothesis.

\vspace{0.1in}
{\it Acknowledgements}. L.O. acknowledges a partial support of this work 
by Center for Earth Observing and Space Research, George Mason University.
S.M. Kopeikin is thankful to M.-K. Fujimoto and other members of National
Astronomical Observatory of Japan (Mitaka, Tokyo), where this
work was initiated, for long-term constant support, and is
pleasured to acknowledge the hospitality of G. Neugebauer and G. Sch\"afer and
other members of the Institute for Theoretical Physics of the Friedrich
Schiller University of Jena. We are grateful to N. Wex who carefully read the
manuscript and made a number of valuable comments which helped to improve the
presentation of the paper. This work has been partially
supported by the Th\"uringer Ministerium f\"ur Wissenschaft, Forschung und 
Kultur grant No B501-96060 and Max-Planck-Gesselschaft grant No 02160/361.

%\newpage

\begin{figure}
\centerline{\psfig{figure=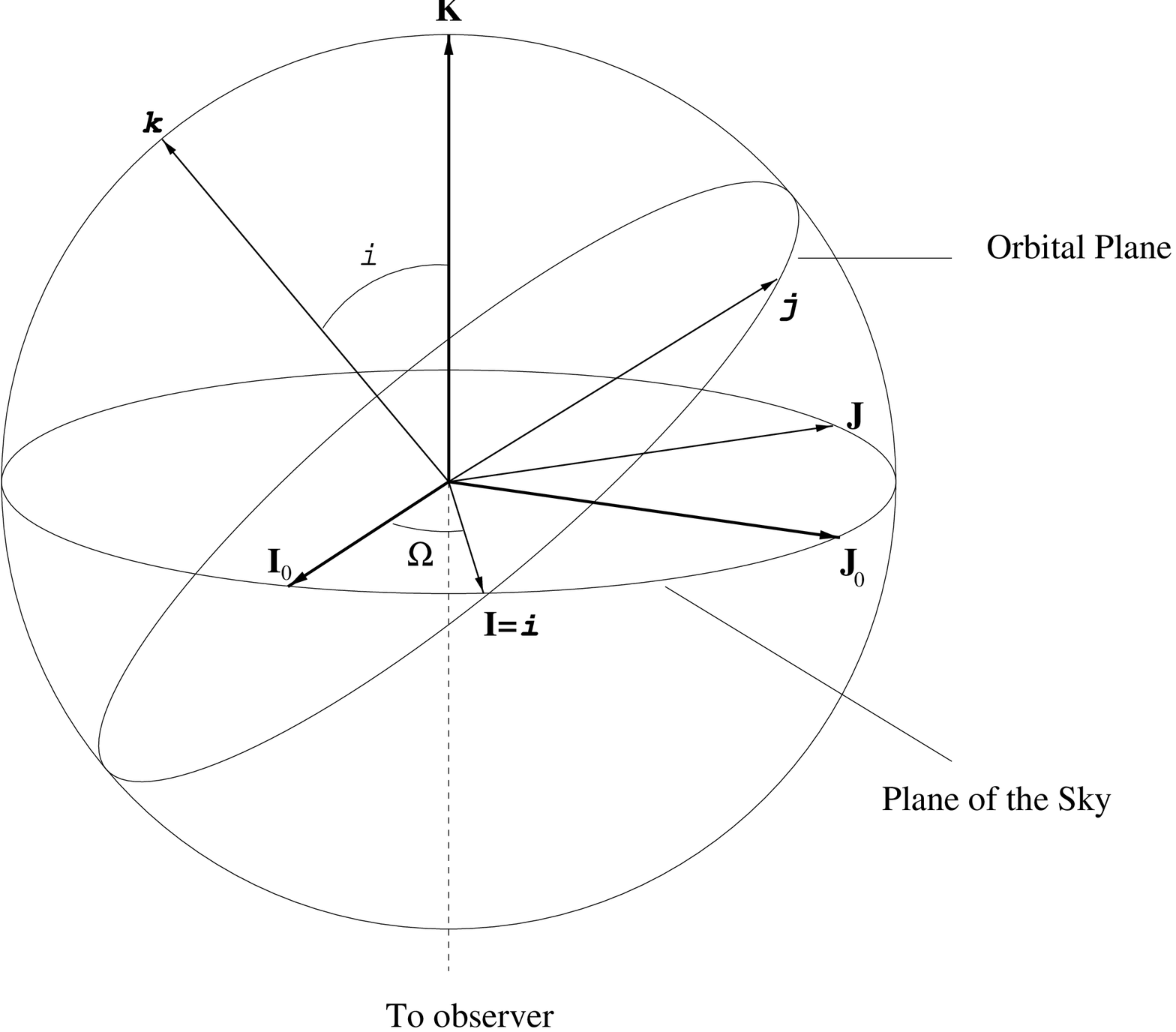,angle=270,width=18cm}}
\vspace{2cm}
\caption{Angles and orientation conventions relating the orbit of the binary
system to the observer's coordinate system and the line of sight. The orbital
plane is inclined at angle $i$ with respect to the plane of the sky. The angle
$\Omega$ is the longitude of the ascending node of the orbital plane.}
\label{Fig.1}
\end{figure}

\begin{thebibliography}{99}
\bibitem{} Armstrong, J. T., Mozurkewich, D., Rickard, L. J, Hutter, D. J.,
Benson, J. A., Bowers, P. F., Elias, N. M. II, Hummel, C. A., Johnston, K. J., 
Buscher, D. F., Clark, J. H. III, Ha, L., Ling, L.-C., White, N. M., \&
Simon, R. S., 1998, ApJ, {\bf 496}, 550

\bibitem{} Arzoumanian, Z., Joshi, K., Rasio, F. A., Thorsett, S. E., 1996,
"Pulsars: problems and progress",
Astronomical Society of the Pacific Conference Series,
Vol. 105; Proceedings of the 160th colloquium of the International Astronomical Union
held in Sydney; Australia; 8-12 January 1996; San Francisco: Astronomical Society of
the Pacific (ASP); 1996; edited by S. Johnston, M.A. Walker, and M. Bailes., 
p.525

\bibitem{}  Brumberg, V. A., 1972, Relativistic Celestial Mechanics. Nauka:
Moscow ({\it in Russian})

\bibitem{}  Brumberg, V. A., 1991, Essential Relativistic Celestial
Mechanics. Bristol: Adam Hilger

\bibitem{}  Brumberg, V. A., \& Kopeikin, S. M., 1989a, ``Relativistic theory
of celestial reference frames'', in: Reference Frames in Astronomy and
Geophysics, eds. J. Kovalevsky, I. I. Mueller, \& B. Kolaczek, Kluwer:
Dordrecht, p. 115 

\bibitem{}  Brumberg, V. A., \& Kopeikin, S. M., 1989b, Nuovo Cim., 
{\bf 103B}, 63

\bibitem{}  Brumberg, V. A., \& Kopeikin, S. M., 1990, Cel. Mech, {\bf 48}, 23

\bibitem{}  Brumberg, V. A., Novikov, I. D., Shakura, N. I., \& Zel'dovich,
Ya. B., 1975, Sov. Astron. Lett., {\bf 1}, 2

\bibitem{}  Cochran, W., 1996, "Precise Measurement of Stellar Radial
Velocities", In: Proc. Workshop on High Resolution Data Processing, 
Eds., M. Iye, T. Takata, and J. Wampler, 1996, SUBARU Telescope Techical 
Report, NAOJ, No 55, p. 30

\bibitem{}  Damour, T., \& Deruelle, N., 1985, Ann. de l'Inst. H. Poincar\'e,
Phys. Th\'eorique, A{\bf 43}, 107

\bibitem{}  Damour, T. \& Deruelle, N., 1986b, Ann. de l'Inst. H. Poincar\'e,
Phys. Th\'eorique, A{\bf 44}, 263

\bibitem{}  Damour, T., Soffel, M., \& Xu, C., 1991, Phys. Rev. D, {\bf 43},
3273

\bibitem{}  Damour, T., Soffel, M., \& Xu, C., 1992, Phys. Rev. D, {\bf 45},
1017

\bibitem{}  Damour, T., Soffel, M., \& Xu, C., 1993, Phys. Rev. D, {\bf 47},
3124

\bibitem{}  Damour, T., \& Taylor, J. H., 1992, Phys. Rev. D, {\bf 45}, 1840

\bibitem{}  Doroshenko, O. V., \& Kopeikin, S. M., 1990, Sov. Astron., {\bf 34%
}, 496

\bibitem{}  Doroshenko, O. V., \& Kopeikin, S. M., 1995, MNRAS, {\bf 274},
1029

\bibitem{}  Guinan, E. F., \& Maloney, D., 1985, AJ,{\bf \ 90}, 1519

\bibitem{}  Hajian, A.R. 1998, private communication

\bibitem{}  IERS Annual Reports 1988 - 95, Paris, France

\bibitem{} Khaliullin, K. F., 1985, ApJ, {\bf 299}, 668

\bibitem{}  Klioner, S. A., \& Kopeikin, S. M., 1994, ApJ, {\bf 427}, 951

\bibitem{}  Kopeikin, S. M., 1988, Cel. Mech., {\bf 44}, 87

\bibitem{}  Kopeikin, S. M., 1994, ApJ, {\bf 434}, L67 

\bibitem{}  Kopeikin, S. M., 1996, ApJ, {\bf 467}, L93

\bibitem{}  Kopeikin, S. M., 1997, J. Math. Phys., {\bf 38}, 2587

\bibitem{}  Kopeikin, S. M., Sch{\"a}fer, G., Gwinn, C. R., \& Eubanks, T. M.,
1998, preprint gr-qc/9811003, accepted by Phys. Rev. D

\bibitem{}  Kopeikin, S. M., \& Ozernoy, L. M., 1996, Bull. Amer. Astron. Soc., 
{\bf 188},  87.03

\bibitem{}  Landau, L. D., \& Lifshitz, E. M., 1951, The classical Theory of
Fields. Addison-Wesley: Cambridge 

\bibitem{}  Ozernoy, L. M., 1997a, Bull. Amer. Phys. Soc. {\bf 42}, 1111 

\bibitem{}  Ozernoy, L. M., 1997b,  MNRAS, {\bf 291}, L63

\bibitem{} Robertson, H. P., 1938, Ann. Math., {\bf 39}, 101

\bibitem{}  Semeniuk, I., \& Paczynski, B., 1968, Acta. Astr., {\bf 18}, 33

\bibitem{} Shakura, N. I., 1985, Sov. Astr. Lett., {\bf 11}, 224

\bibitem{}  Shapiro, I. I., 1964, Phys. Rev. Lett., {\bf 13}, 789

\bibitem{}  Shore, S. N., 1994, Observations and Physical Processes in
Binary Stars, In: Interacting Binaries, Eds., S. N. Shore, M. Livio, E. P.
J. van den Heuvel, Berlin: Springer - Verlag, p. 1 

\bibitem{ }  Standish, E. M., 1982, A\&A, {\bf 114}, 297

\bibitem{}  Standish, E. M., 1993, AJ, {\bf 105}, 2000

\bibitem{}  Taylor, J. H., 1992, Phil. Trans. R. Soc. Lond. A., {\bf 341},
117

\bibitem{}  Taylor, J. H., 1994, Rev. Mod. Phys., {\bf 66}, 711

\bibitem{}  Taylor, J. H., \& Weisberg, J. M., 1989, ApJ, {\bf 345}, 434

\bibitem{}  Tolman, R. C., 1934, Relativity, Thermodynamics and Cosmology.
Oxford Clarendon Press: London

\bibitem{}  Valenti, J.A., Butler, R. P., \& Marcy, G. W., 1995, PASP, {\bf %
107}, 966

\bibitem{}  Weinberg, S., 1972, Gravitation and Cosmology, John Wiley \& Sons: 
New York

\bibitem{} Wex, N., Johnston, S., Manchester, R. N., Lyne, A. G., 
Stappers, B. W., \& Bailes, M., 1998, MNRAS, {\bf 298}, 997 

\end{thebibliography}
\end{document}